\documentclass{article}

\usepackage[final,nonatbib]{neurips_2020}

\usepackage[utf8]{inputenc} 
\usepackage[T1]{fontenc}    
\usepackage{hyperref}       
\usepackage{url}            
\usepackage{booktabs}       
\usepackage{amsfonts}       
\usepackage{nicefrac}       
\usepackage{microtype}      

\usepackage{amsmath, amssymb, amsfonts, amsthm}
\usepackage{graphicx}
\usepackage{url}
\usepackage{bm}
\usepackage{subfigure}
\usepackage{graphicx}
\usepackage{mathabx}
\usepackage{multirow}
\usepackage{setspace}
\usepackage[separate-uncertainty = true,multi-part-units=single]{siunitx}
\usepackage{biocon}
\usepackage{booktabs}
\usepackage{outlines}
\usepackage{textcomp}
\usepackage{tabularx}
\usepackage{makecell}
\usepackage{rotating}
\usepackage{lscape} %
\usepackage{longtable} %
\usepackage[roman]{parnotes}
\usepackage[version=4]{mhchem}
\usepackage{contour} 
    \contourlength{0.8pt}
\usepackage{verbatim}
\usepackage{booktabs}
\usepackage{multirow}
\usepackage[table,xcdraw,dvipsnames]{xcolor}
\usepackage{caption}
\usepackage{float} 
\usepackage{array} 
\usepackage{longtable}
\usepackage{csvsimple}
\usepackage{pdfpages}
\usepackage{longtable}
\usepackage{authblk}

\newcommand{\concat}{%
  \mathbin{{+}\mspace{-8mu}{+}}%
}

\title{Explainable multiple abnormality classification of chest CT volumes}

\author[a]{Rachel Lea Draelos, M.D.,  Ph.D.}
\author[b]{Lawrence Carin, Ph.D.}
\affil[a]{Corresponding author. rlb61@duke.edu. Duke University Computer Science}
\affil[b]{lawrence.carin@kaust.edu.sa. Duke University Electrical and Computer Engineering}

\begin{document}

\maketitle 


\begin{abstract}
Understanding model predictions is critical in healthcare, to facilitate rapid verification of model correctness and to guard against use of models that exploit confounding variables. We introduce the challenging new task of explainable multiple abnormality classification in volumetric medical images, in which a model must indicate the regions used to predict each abnormality. To solve this task, we propose a multiple instance learning convolutional neural network, AxialNet, that allows identification of top slices for each abnormality. Next we incorporate HiResCAM, an attention mechanism, to identify sub-slice regions. We prove that for AxialNet, HiResCAM explanations are guaranteed to reflect the locations the model used, unlike Grad-CAM which sometimes highlights irrelevant locations. Armed with a model that produces faithful explanations, we then aim to improve the model's learning through a novel mask loss that leverages HiResCAM and 3D allowed regions to encourage the model to predict abnormalities based only on the organs in which those abnormalities appear. The 3D allowed regions are obtained automatically through a new approach, PARTITION, that combines location information extracted from radiology reports with organ segmentation maps obtained through morphological image processing. Overall, we propose the first model for explainable multi-abnormality prediction in volumetric medical images, and then use the mask loss to achieve a 33\% improvement in organ localization of multiple abnormalities in the RAD-ChestCT data set of 36,316 scans, representing the state of the art. This work advances the clinical applicability of multiple abnormality modeling in chest CT volumes.
\end{abstract}


\section{Introduction}

Automated interpretation of medical images with machine learning has the potential to revolutionize the field of radiology. However, machine learning systems have not yet been adopted on a large scale in clinical practice \cite{alexander2020market}. One barrier to adoption is trust \cite{lim2019building, ribeiro2016should}. Most medical imaging models are based on convolutional neural networks (CNNs), which are ``black box'' models unless additional steps are taken to improve explainability \cite{ancona2019XAIBook}. 

Visual explanation methods in computer vision indicate which regions of an input image contribute to a model's predictions \cite{ancona2019XAIBook}. Explainability is critical in medical imaging to detect situations where a model leverages aspects of the data that are correlated with an outcome but inappropriate for prediction. Zech \emph{et al.} \cite{zech2018variable} used class activation mapping \cite{zhou2016CAM} to reveal that a CNN trained to predict pneumonia from chest radiographs leveraged non-medical features to make the pneumonia prediction. Specifically, the model identified differences in metal tokens, postprocessing, and compression artefacts which revealed the hospital system of origin, a highly effective indicator of pneumonia risk due to differing frequencies of pneumonia among the patient populations. Furthermore, a ``model'' consisting of sorting the radiographs by hospital system achieved an AUROC of 0.861, illustrating that high performance alone does not guarantee that a model is faithful to expected medical reasoning. This behavior is analogous to natural image classifiers detecting boats via water, trains via rails, or horses via a copyright watermark \cite{lapuschkin2016analyzing, lapuschkin2019unmasking}. It is thus critical to seek insight into how models make their predictions.

In this work, we consider explainable multiple abnormality classification in volumetric medical images, a task that to the best of our knowledge has not yet been explored in the literature. The four new contributions of this paper are summarized below. Our first two contributions create an initial solution:

\begin{itemize}
    \item We propose AxialNet, a multiple instance learning CNN that explains its predictions by identifying the axial slices that contribute most towards identification of each abnormality;
    \item HiResCAM is a previously published method with faithfulness guarantees. To obtain finer-grained, sub-slice predictions, we incorporate HiResCAM explanations into AxialNet, and include a new proof that shows for the first time that HiResCAM's faithfulness guarantee also holds for the class of multiple instance learning models AxialNet represents.
\end{itemize}
    
Our next two contributions aim to improve AxialNet's learning and thus improve the ``reasoning'' behind its explanations:
\begin{itemize}
    \item We develop a novel mask loss that leverages HiResCAM and 3D allowed regions to encourage AxialNet to predict abnormalities from only the organs in which they are found;
    \item In order to obtain the 3D allowed regions efficiently, we propose PARTITION, a method to obtain pixel-level, abnormality-specific allowed regions without any manual labeling, by combining location information extracted from radiology reports with organ segmentation maps obtained via morphological image processing.
\end{itemize}

Overall, we present an initial method for the new task of explainable multiple abnormality classification in volumetric medical images, then improve upon our solution to achieve state-of-the-art performance. This work represents a step towards machine learning systems that may accelerate the radiology workflow and contribute to improved detection and monitoring of disease.

\section{Related Work}

\subsection{Multiple abnormality modeling in CT volumes}

Computed tomography (CT) scans are used to diagnose and monitor numerous conditions, including cancer \cite{shieh2017lowdose}, injuries \cite{oikonomou2011ctbluntchesttrauma}, and lung disease \cite{crossley2018ctemphysema, raju2017chestpictoral}. Due to the challenging and time-consuming nature of CT interpretation, there has been substantial interest in developing machine learning models to analyze CT scans. Almost all prior work in CT classification has focused on one class of abnormalities at a time, such as interstitial lung disease \cite{bermejo2020classification, wang2019weakly, gao2016multi, gao2018holistic, walsh2018deep, christodoulidis2016multisource, anthimopoulos2016lung}, lung cancer \cite{ardila2019end}, pneumothorax \cite{li2019deep}, or emphysema \cite{humphries2020deep}. The only model developed to predict multiple diverse abnormalities simultaneously from one CT volume is CT-Net, which was trained and evaluated on the RAD-ChestCT data set of 36,316 CT volumes \cite{Draelos, ZenodoRADChestCT}. While CT-Net achieves high performance, its final representation is not interpretable due to an intermediate convolution step over the feature dimension that disrupts the spatial relationship between the input volume and the low-dimensional representation. We propose a new model, AxialNet, which outperforms CT-Net while providing explainability through identification of the slices that contribute most to each abnormality prediction.

\subsection{Class-specific explanation methods for CNNs}

Slice-level explanations are useful, but obtaining finer-grained explanations remains important for gaining additional insight into a model's behavior. One straightforward way to obtain abnormality-specific fine-grained explanations is via the gradient-based methods Class Activation Mapping (CAM) \cite{zhou2016CAM} and Grad-CAM \cite{selvaraju2017grad}, which are commonly used explanation approaches in medical imaging \cite{lee2019explainable, panwar2020deep, baltruschat2019comparison, shen2018dynamic, pasa2019efficient, lu2019deep}. Unfortunately, CAM can only be used for CNNs that end in global average pooling followed by one fully connected layer, which does not apply to AxialNet. Grad-CAM is a generalization of CAM that loosens CAM's architecture requirements; however, recent work has demonstrated that Grad-CAM can be misleading, and sometimes highlights irrelevant locations that the model did not actually use for prediction \cite{draelos2020usehirescam}. HiResCAM is a new explanation method that is applicable to wider range of CNN architectures than CAM, while remaining provably guaranteed to accurately highlight the locations these models used \cite{draelos2020usehirescam}. In this work, we prove that HiResCAM's location faithfulness is additionally guaranteed for the multiple instance learning architecture of the AxialNet model. We then apply HiResCAM to AxialNet in order to obtain sub-slice explanations.

\subsection{Abnormality segmentation in CT volumes}

AxialNet and HiResCAM provide an initial solution to the task of explainable abnormality prediction. In an effort to improve the concepts AxialNet learns, we propose a mask loss that encourages the model to identify abnormalities from within relevant anatomical structures. The mask loss is mathematically related to the segmentation cross entropy objective commonly used when training abnormality segmentation models \cite{jadon2020survey}. Like a segmentation loss, our mask loss does rely on abnormality-specific allowed regions. However, unlike a segmentation loss, the mask loss is intended to enhance a classification model rather than train an abnormality segmentation model; it is calculated in a low-dimensional space for computational feasibility, rather than the input space; and it relies on automatically generated allowed regions based on organs rather than manually obtained abnormality segmentation maps, so that over 80 abnormalities across 36,316 CT volumes can be considered rather than 1-2 abnormalities across a few hundred CT volumes.

\subsection{Organ segmentation in CT volumes}

The abnormality-specific allowed regions for our mask loss are based on organ segmentations, and are obtained using PARTITION, which is the first reported approach to provide abnormality-specific allowed regions automatically. The organ segmentation in PARTITION uses morphological image processing. We choose an unsupervised morphological image processing approach over a supervised machine learning approach for several reasons. 

First, the unsupervised approach for organ segmentation requires no manually created segmentation maps. RAD-ChestCT \cite{Draelos} includes whole-volume abnormality labels, but no segmentation ground truth, so training a machine-learning-based multi-organ segmentation model \cite{dong2019automatic, chen2020automatic, zhou2020automatic, peng2020method} on RAD-ChestCT would require manually circumscribing the relevant anatomical structures on hundreds or thousands of slices, representing months of effort by a domain expert.

Second, the two largest organs we were interested in segmenting - the right and left lungs - can be segmented with excellent performance using morphological image processing, because they are large contiguous regions of mostly black pixels. In fact, morphological image processing yields such good results for lung segmentation that it was used to create the training set lung segmentation ground truth in the CT-ORG data set \cite{CTorgDataset,CTorgAcknowledge,CTorgTCIA}. 

Finally, training a machine learning segmentation model on another dataset and then deploying it on RAD-ChestCT would be unlikely to result in satisfactory performance on RAD-ChestCT due to significant differences in the data distributions \cite{dou2019unsupervised}. RAD-ChestCT includes only non-contrast chest CT scans, while CT-ORG includes contrast, non-contrast, abdominal, and full body CT scans, and the AAPM Thoracic Auto-segmentation Challenge 2017 data set \cite{segchallenge2017,yang2018autosegmentation} includes only contrast scans. RAD-ChestCT includes a mixture of mild and severe lung diseases, and 95\% of scans have a slice thickness $<$0.625 mm; AAPM excludes cases with collapsed lungs from extensive disease, and uses scans with slice spacing of 1 mm, 2.5 mm, or 3 mm. The anatomical regions of interest also differ. For our application, we required right lung, left lung, and heart/great vessels segmentations; CT-ORG includes labels for lung but not the heart or great vessels, while AAPM includes labels for lungs and heart but not the great vessels. 
For these reasons, we decided that domain adaptation from CT-ORG and/or AAPM models to RAD-ChestCT was best left to future work, and we pursued an unsupervised organ segmentation approach that could be optimized for the RAD-ChestCT dataset.

\section{Methods}

\begin{figure}
\begin{center}
    \includegraphics[scale=0.34]{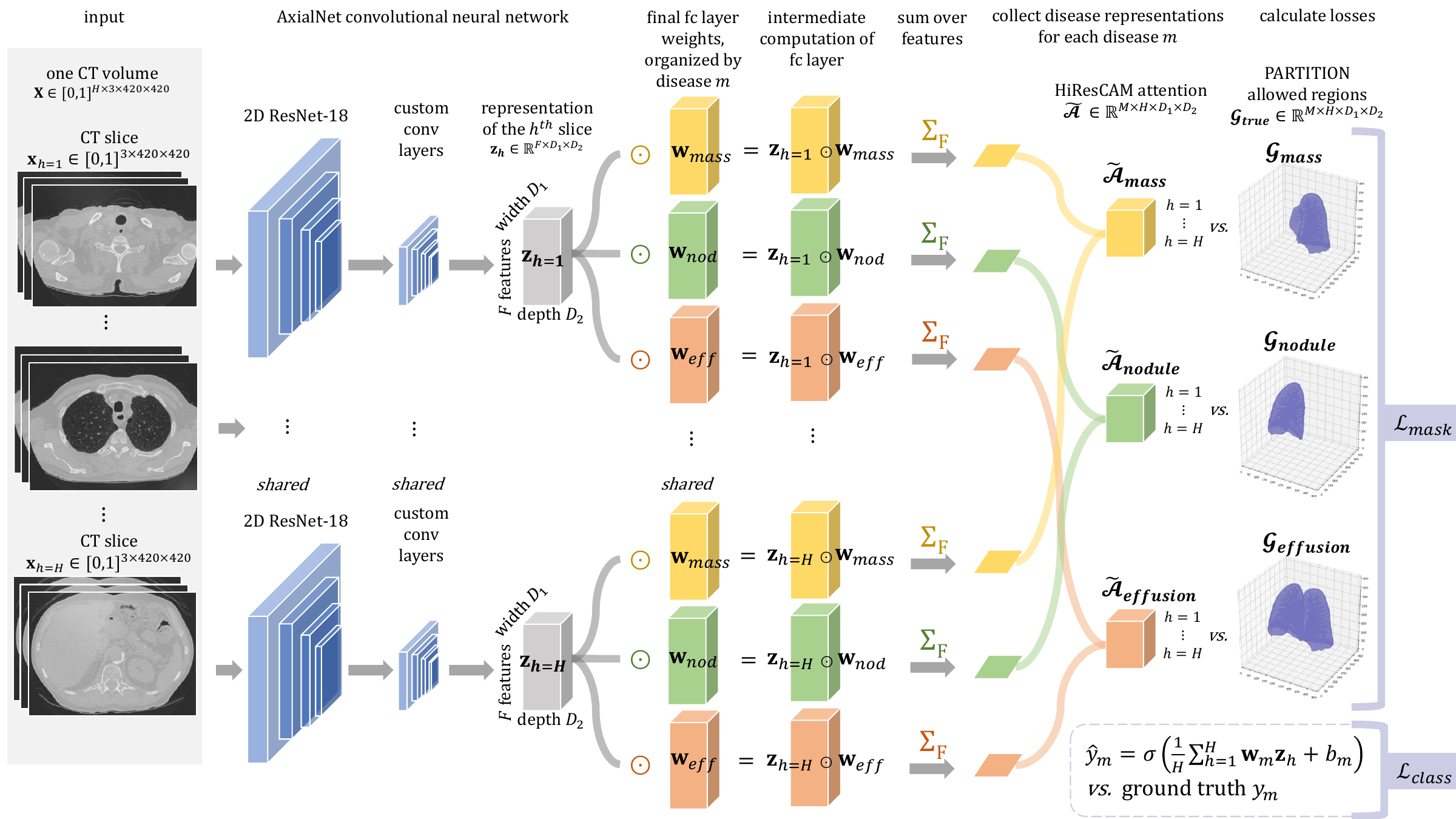}
    \caption{The proposed AxialNet model including a mask loss calculated using HiResCAM attention and PARTITION allowed regions. In the initial layers of AxialNet, a low-dimensional representation $\mathbf{z}_h$ is obtained for each slice input $h=1,...,H$ via 2D convolutions. A final fully connected layer produces $M$ abnormality scores per slice. The slice scores are averaged for each abnormality to produce the overall score used in the classification loss (lower right). A HiResCAM explanation is calculated for each abnormality as the element-wise product of the overall scan representation $\mathbf{Z}$ with the gradient of the abnormality score with respect to $\mathbf{Z}$, summed over the feature dimension. In AxialNet, this gradient is directly proportional to the fully connected layer weights for the relevant abnormality. The mask loss we propose is computed using the HiResCAM explanations and PARTITION allowed regions $\mathcal{G}_{true}$ to encourage the model to only increase the abnormality score using organs in which that abnormality is found. Here, the allowed organs are the left lung for mass, the right lung for nodule, and both lungs for pleural effusion. Best viewed in color.}
    \label{fig:model}
\end{center}
\end{figure}

\subsection{Problem setup}

Consider a dataset of $N$ examples $\{\textbf{X},\textbf{y}\}_{i=1}^N$ where $\textbf{y}_i\in\{0,1\}^M$ is a binary vector corresponding to the presence/absence of $M$ abnormalities and $\textbf{X}_i \in [0,1]^{405 \times 420 \times 420}$ is a CT volume. We wish to predict, in an explainable manner, the labels $\hat{\mathbf{y}}_i$ given a CT volume $\mathbf{X}_i$, meaning that for any abnormality $m\in\{1,2,\dots,M\}$ a physician can query the model to obtain a visualization that highlights the sub-regions of the volume that the model used to predict that abnormality. Figure \ref{fig:model} provides an overview of our solution to this task.

\subsection{AxialNet architecture}

Radiologists often view CT scans as a stack of axial slices \cite{whiting2015computedbasicprinciples}, which form horizontal planes through an upright patient. Motivated by this practice, we propose AxialNet, a multiple instance learning architecture that treats each CT scan as a ``bag of slices'' and produces whole-volume scores by averaging per-slice scores, thus enabling direct determination of which axial slices contributed the most to prediction of each abnormality. 

In the first part of AxialNet, a 2D CNN is applied to each CT slice. The CNN's parameters are shared across slices (Figure \ref{fig:model}), an approach that has been previously successful in CT analysis \cite{Draelos, li2020COVIDresnet, saab2019doubly} and has the benefit of reducing the number of model parameters while incorporating the reasonable assumption that the same features may be seen at multiple levels in a CT volume. The 2D CNN applied to each CT slice is implemented using a ResNet-18, followed by custom convolutional layers that further reduce the size of the representation and result in improved performance.
Implementation details: The pretrained ResNet-18 requires 3-channel input. One solution would be to triplicate each CT slice, but we were unable to pursue this due to memory limits. Thus, we reshaped each CT volume to $\textbf{X}_i \in [0,1]^{135 \times 3 \times 420 \times 420}$, grouping together three adjacent axial slices. We use the term ``slices'' in the paper but our experiments more precisely use ``slice groups.'' $H$ refers to the input to the 2D CNN, which is a slice (if one slice is triplicated) or a slice group (if three slices are considered together).
The low-dimensional representation produced by the 2D CNN for the $h^{th}$ slice is termed $\mathbf{z}_h \in \mathbb{R}^{F \times D_1 \times D_2}$ with $F$ features, width $D_1$, and depth $D_2$. 

An abnormality score vector $\mathbf{c}_h \in \mathbb{R}^M$ is then produced from each slice representation $\mathbf{z}_h$ using a fully-connected layer with parameters shared across all CT slices: $\mathbf{c}_h = \mathbf{W}\mathbf{z}_h + \mathbf{b}$, with $\mathbf{W} \in \mathbb{R}^{M \times F D_1 D_2}$ and $\mathbf{b} \in \mathbb{R}^M$. Since AxialNet makes predictions for all $M$ abnormalities and all $H$ slices at once, it produces a matrix of all per-slice abnormality scores $\mathbf{C} \in \mathbb{R}^{M \times H}$. This matrix $\mathbf{C}$ provides basic explainability because it illustrates the quantitative contribution of each axial slice to each abnormality score, enabling identification of the top slices for each abnormality prediction.

Next, for each abnormality $m$ AxialNet averages together the per-slice scores to produce a whole-volume abnormality score $s_m$:
\begin{align}
    s_m = \frac{1}{H} \sum_{h=1}^H \mathbf{w}_m \mathbf{z}_h + b_m = \frac{1}{H} \sum_{h=1}^H c_{mh}. \label{eqn:abnscoressm}
\end{align}
Above, one row vector $\mathbf{w}_m \in \mathbb{R}^{1 \times F D_1 D_2}$ of the weight matrix $\mathbf{W}$ corresponds to one abnormality, meaning that the expression $\mathbf{w}_m \mathbf{z}_h + b_m$ produces the scalar score $c_{mh}$ for the $m^{th}$ abnormality and $h^{th}$ slice.

The whole-volume predicted probability $\hat{y}_m$ for the $m^{th}$ abnormality is calculated from the whole-volume score $s_m$ using the sigmoid function $\hat{y}_m = \sigma(s_m) = \frac{1}{1+e^{-s_m}}$. 

For each observation we optimize the expected multilabel cross entropy objective, requiring only whole-volume abnormality labels:
\begin{align}
    \mathcal{L}_{class} = -\frac{1}{M} \sum_{m=1}^{M} [y_m \log \hat{y}_m + (1-y_m)\log (1-\hat{y}_m)]. \label{eqn:crossentropy}
\end{align}

\subsection{Applying HiResCAM for abnormality-specific attention}  \label{section:hirescam}

AxialNet alone can provide explainability through identification of top slices. However, AxialNet does not provide granular sub-slice explanations. In order to obtain these finer-grained explanations, we apply High-Resolution Class Activation Mapping (HiResCAM) \cite{draelos2020usehirescam}, a recently proposed visual explanation method. Recall that $s_m$ is the model's score for abnormality $m$ before the sigmoid function (equation \ref{eqn:abnscoressm}). To obtain a HiResCAM explanation for abnormality $m$, we first compute the gradient of $s_m$ with respect to a collection of convolutional feature maps $\mathbf{A}= \{ \mathbf{A}^f \}_{f=1}^F$. For volumetric data, this gradient $\frac{\partial s_m}{\partial \mathbf{A}}$ is 4-dimensional $[F, H, D_1, D_2]$. The HiResCAM explanation is an element-wise multiplication of this gradient with the feature maps themselves:
\begin{align} \label{eqn:hirescam}
    \mathcal{\tilde{A}}_m^{\rm{HiResCAM}} = \sum_{f=1}^{F} \frac{\partial s_m}{\partial \mathbf{A}^f} \odot \mathbf{A}^f.
\end{align}

\subsection{HiResCAM provably highlights relevant locations for AxialNet models} \label{sec:hires-prove}

For any CNN consisting of convolutional layers followed by a single fully connected layer, HiResCAM was previously proven to highlight locations the model used when applied at the last convolutional layer \cite{draelos2020usehirescam}. In this section, we prove that this location faithfulness guarantee also holds for CNNs that additionally include a final multiple instance learning averaging layer, \textit{i.e.} models with the general structure of AxialNet. We thus expand the class of CNNs for which HiResCAM's faithfulness guarantee holds.

\begin{proof}

We apply HiResCAM at the last convolutional layer of AxialNet; this layer produces a low-dimensional representation of the entire CT scan termed $\mathbf{Z}$. We thus replace $\mathbf{A} = \{ \mathbf{A}^f \}_{f=1}^{F}$ in equation \ref{eqn:hirescam} with the feature maps $\mathbf{Z}=\{ \mathbf{Z}^f \}_{f=1}^F$ to obtain:
\begin{align}
    \mathcal{\tilde{A}}_m^{\rm{HiResCAM}} = \sum_{f=1}^{F} \frac{\partial s_m}{\partial \mathbf{Z}^f} \odot \mathbf{Z}^f. \label{eqn:hirescamzfora}
\end{align}
To calculate $\frac{\partial s_m}{\partial \mathbf{Z}}$, the gradient of the abnormality score $s_m$ with respect to $\mathbf{Z}$, we must use an expression for $s_m$. However, the previous expression for the abnormality score $s_m = \frac{1}{H} \sum_{h=1}^H \mathbf{w}_m \mathbf{z}_h + b_m$ (equation \ref{eqn:abnscoressm}) expressed the score in terms of the slice representations $\mathbf{z}_h$ rather than $\mathbf{Z}$ overall.

We can rewrite the score $s_m$ in terms of $\mathbf{Z}$ overall via two concatenations. First, define $\mathbf{Z}$ as the vector resulting from concatenation of all the flattened $\mathbf{z}_h$ representations, $\mathbf{Z} = \mathbf{z}_1 \concat \mathbf{z}_2 \concat ... \concat \mathbf{z}_H$, with flattened $\mathbf{Z} \in \mathbb{R}^{H F D_1 D_2 \times 1}$. Next, define $\mathbf{w}_m^{cat}$ as the vector resulting from concatenation of the $m^{th}$-abnormality-specific weights $\mathbf{w}_m$ with themselves $H$ times: $\mathbf{w}_m^{cat} = \mathbf{w}_m \concat \mathbf{w}_m \concat ... \concat \mathbf{w}_m$, where $\mathbf{w}_m \in \mathbb{R}^{1 \times F D_1 D_2}$ and $\mathbf{w}_m^{cat} \in \mathbb{R}^{1 \times H F D_1 D_2}$. Then an alternative expression for the whole volume abnormality score $s_m$ is:
\begin{align}
   s_m = \frac{1}{H} \mathbf{w}_m^{cat} \mathbf{Z} + b_m.    \label{eqn:scoreAlternative}
\end{align}
Note that the $\frac{1}{H}$ fraction is only applied to the $\mathbf{w}_m^{cat} \mathbf{Z}$ term because $\frac{1}{H} \times H b_m = b_m$.

The gradient of the abnormality score $s_m$ with respect to $\mathbf{Z}$ can then be calculated as:
\begin{align}
    \frac{\partial s_m}{\partial \mathbf{Z}} &=  \frac{\partial}{\partial \mathbf{Z}} \left( \frac{1}{H} \mathbf{w}_m^{cat} \mathbf{Z} + b_m \right) \\
    &= \frac{1}{H} \mathbf{w}_m^{cat}. \label{eqn:gradientresult}
\end{align}
Substituting equation \ref{eqn:gradientresult} for $\frac{\partial s_m}{\partial \mathbf{Z}}$ into equation \ref{eqn:hirescamzfora}, we obtain
\begin{align}
    \mathcal{\tilde{A}}_m^{\rm{HiResCAM}} = \sum_{f=1}^{F} \frac{1}{H} \mathbf{w}_m^{cat} \odot \mathbf{Z}.   \label{eqn:hiresThisModel}
\end{align}
The element-wise multiplication $\mathbf{w}_m^{cat} \odot \mathbf{Z}$ in the HiResCAM expression is the intermediate computation in calculating $\mathbf{w}_m^{cat} \mathbf{Z}$, which in turn is a direct contributor to the abnormality score $s_m = \frac{1}{H} \mathbf{w}_m^{cat} \mathbf{Z} + b_m$. 

Therefore, the large positive elements of the HiResCAM explanation $\mathcal{\tilde{A}}_m^{\rm{HiResCAM}}$ that show up as abnormality-relevant image locations correspond to locations which directly increase the abnormality score. Similarly, negative elements of the HiResCAM explanation are direct contributors to a lower abnormality score. Thus, for any CNN consisting of convolutional layers, one fully connected layer, and a final multiple instance learning averaging layer, HiResCAM explanations at the last convolutional layer can be interpreted as showing exactly which parts of the input CT volume contributed most to each abnormality prediction.

\end{proof}



\subsection{HiResCAM versus Grad-CAM} 

HiResCAM is a member of the CAM family of gradient-based explanation methods. Grad-CAM is another member of this family, and is an explanation method that is familiar to many medical imaging researchers. Unfortunately, Grad-CAM does not have a location faithfulness guarantee, which means that Grad-CAM sometimes produces misleading explanations that highlight irrelevant locations \cite{draelos2020usehirescam}. To demonstrate the effect of these misleading explanations in a medical imaging context, we compare Grad-CAM to HiResCAM for the task of explainable multiple abnormality prediction in CT volumes. In order to apply Grad-CAM to CT images, we must extend Grad-CAM from 2D to 3D. 

\subsubsection{Extending Grad-CAM to 3D Data}

Similar to HiResCAM, Grad-CAM is applied at the output of a convolutional layer. The first step is the same as HiResCAM: calculate  $\frac{\partial s_m}{\partial \mathbf{A}}$, the gradient of $s_m$ with respect to a collection of feature maps $\mathbf{A}= \{ \mathbf{A}^f \}_{f=1}^F$. The next step differs: we calculate a vector of importance weights \cite{selvaraju2017grad} $\boldsymbol{\alpha}_m \in \mathbb{R}^F$ that will be used to re-weight each corresponding feature map $\mathbf{A}^f$. For 3D data, the importance weights are obtained by global average pooling the gradient over the height, width, and depth dimensions:
\begin{align}
    \alpha_m^f = \frac{1}{H D_1 D_2} \sum_{h=1}^H \sum_{d_1=1}^{D_1} \sum_{d_2=1}^{D_2} \frac{\partial s_m}{\partial \mathbf{A}^f_{h d_1 d_2}}. \label{eqn:importanceweights}
\end{align}
The importance weights suggest which features are most relevant to this particular abnormality throughout the volume overall. The final Grad-CAM explanation is an importance-weighted combination of the feature maps $\mathbf{A}^f$:
\begin{align} \label{eqn:vanillagradcam}
    \mathcal{\tilde{A}}_m^{\rm{GradCAM}} = \sum_{f=1}^F \alpha_m^f \mathbf{A}^f.
\end{align}

The gradient averaging step that Grad-CAM requires (equation \ref{eqn:importanceweights}) is the reason that Grad-CAM explanations sometimes highlight incorrect locations. For a more detailed discussion of this topic, see \cite{draelos2020usehirescam}.

\subsection{Proposed mask loss} \label{methods:maskloss}

The combination of AxialNet and HiResCAM provides an initial solution to the new task of explainable multiple abnormality classification. The explanations produced are guaranteed to reveal the locations the model used for each abnormality prediction. Initial inspection of the explanations produced by an AxialNet model trained only on a classification loss showed that this model sometimes made predictions using unexpected locations, and thus could be exploiting confounding variables. To encourage the AxialNet model to learn more medically meaningful relationships and thereby produce better explanations, we introduced a mask loss.

Our proposed mask loss drives the model to focus its abnormality-specific attention on 3D allowed regions in which each abnormality appears. Given the model's predicted attention $\mathcal{\tilde{A}} \in \mathbb{R}^{M \times H \times D_1 \times D_2}$ and a binary mask $\mathcal{G}_{true} \in \{0,1\}^{M \times H \times D_1 \times D_2}$ which defines an allowed region for the attention for each abnormality, the mask loss is calculated as follows (using $\tilde{a}_i$ to access all $M H D_1 D_2$ elements of $\mathcal{\tilde{A}}$):
\begin{align}
    \mathcal{L}_{mask} =  - \sum_{i: \mathcal{G}_{true} = 0} \log (1-\tilde{a}_i).
\end{align}
The proposed mask loss is conceptually half of a segmentation loss \cite{jadon2020surveyofloss}, applied in a low-dimensional space. To minimize the mask loss the model must not increase the abnormality score using forbidden regions. Further justification for this choice of mask loss is provided in the Appendix. 

To understand the effect of the mask loss, we train one AxialNet model on the classification loss alone, $\mathcal{L}_{class}$ (equation \ref{eqn:crossentropy}), and another AxialNet model on the overall proposed loss that incorporates both the classification loss and the mask loss: $\mathcal{L}_{total} = \mathcal{L}_{class} + \lambda \mathcal{L}_{mask}$, where $\lambda$ is a hyperparameter. We found $\lambda=\frac{1}{3}$ to be an effective value.

\subsection{PARTITION: Creating an attention ground truth}

\begin{figure}
\begin{center}
    \includegraphics[scale=0.4]{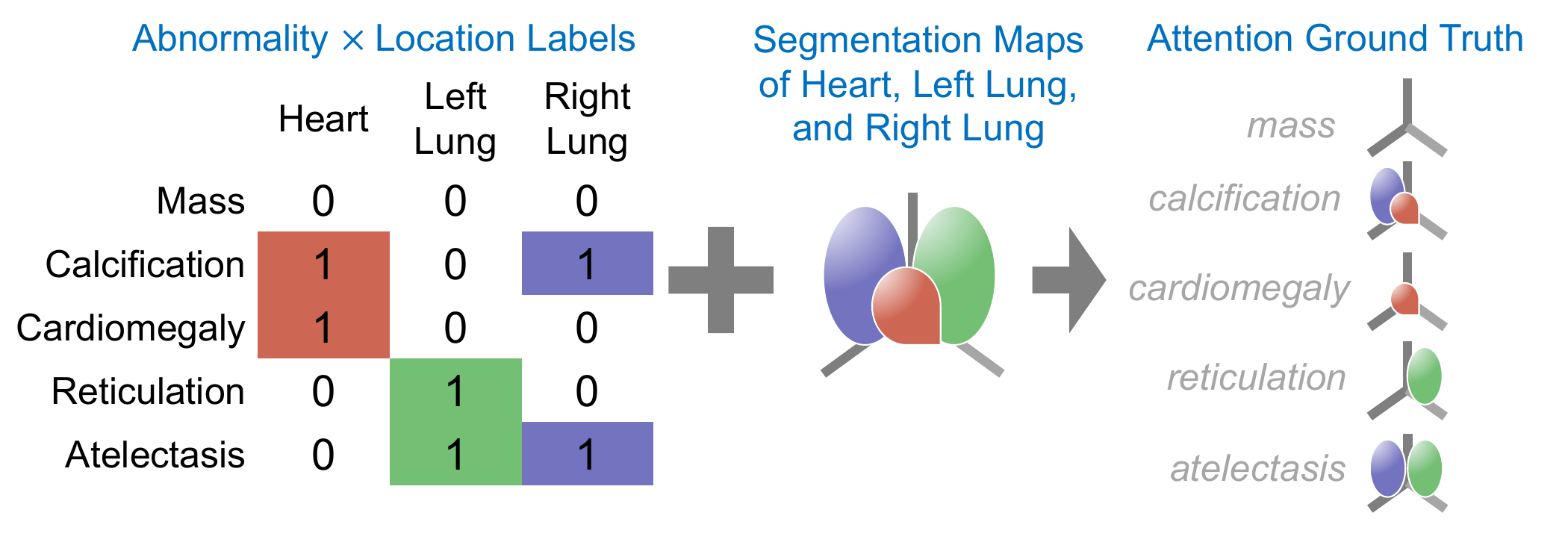}
    \caption{Schematic of the attention ground truth creation via PARTITION. By combining the location $\times$ abnormality labels and the organ segmentation masks, it is possible to define ``allowed regions'' for each abnormality with no manual labor. Best viewed in color.}
    \label{fig:attentiongroundtruth}
\end{center}
\end{figure}

The primary barrier to computing the mask loss is obtaining $\mathcal{G}_{true}$ which specifies the allowed regions for each abnormality. Manually creating $\mathcal{G}_{true}$ would require decades of full-time expert manual labor (36,316 CT volumes $\times$ 405 axial slices $\times$ 80 abnormalities = over one billion annotations; at 1 second per annotation, the full dataset would require 30 years of work). Therefore, we develop PARTITION (``Per Abnormality oRgan masks To guIde aTtentION''), an efficient approach for obtaining $\mathcal{G}_{true}$ automatically, without any manual input. First, we expand the previously described SARLE \cite{Draelos} natural language processing method with location vocabulary, in order to automatically identify the anatomical location of each abnormality described in the free text CT reports. Next, we develop an unsupervised multi-organ segmentation pipeline using morphological image processing to define segmentation maps of the right lung, left lung, and mediastinum in each volume. Combining the location $\times$ abnormality labels with the organ segmentations enables determination of an ``allowed region'' for each abnormality as illustrated in Figure \ref{fig:attentiongroundtruth}. Further details of PARTITION are provided in the Appendix.

\section{Results}

In experiments on RAD-ChestCT, we assess the suitability of AxialNet and HiResCAM for the new task of explainable multiple abnormality prediction in chest CT volumes. We then consider whether the mask loss can encourage the AxialNet model to learn more medically meaningful relationships and thereby produce more appropriate explanations.

\subsection{RAD-ChestCT dataset}

RAD-ChestCT \cite{Draelos, ZenodoRADChestCT} is a data set of 36,316 CT volumes with 83 whole-volume abnormality labels. We focus on $M=80$ labels relevant to the lungs and mediastinum, detailed in the Appendix. To the best of our knowledge, RAD-ChestCT is the only large-scale volumetric medical imaging dataset with multiple diverse abnormality labels. To quantify the effect of the mask loss on organ localization of abnormalities, we train and evaluate AxialNet with and without the mask loss using the full data set of 36,316 volumes ($>5$ terabytes). For architecture comparisons and ablation studies we use a predefined subset \cite{Draelos} of 2,000 training scans and 1,000 validation scans intended for this purpose, as using the full data set would require 1-2 weeks of compute time per comparison.

\subsection{Implementation}

The per-slice CNN in AxialNet consists of a 2D ResNet-18 \cite{he2016deep} pretrained on ImageNet \cite{deng2009imagenet} and refined during training on CTs, followed by custom convolutional layers detailed in the Appendix.  Models were trained using stochastic gradient descent with momentum 0.99 and learning rate $10^{-3}$. Whole-dataset models were trained on an NVIDIA Tesla V100 GPU with 32 GiB of memory. All models are implemented in PyTorch. Code will be made publicly available upon publication\footnote{\href{https://github.com/rachellea}{https://github.com/rachellea}}.

\subsection{OrganIoU: a metric for organ localization of abnormalities}

The accuracy of any explanation method for a particular class of models must be proven mathematically; it is impossible to measure using experiments \cite{draelos2020usehirescam}. However, it \textit{is} possible to experimentally evaluate how well a particular model yields desired behavior, such as localizing abnormalities to the correct organs. 

To gain insight into the behavior of particular trained AxialNet models - specifically, to measure the models' organ localization of abnormalities - we propose OrganIoU, a metric that equals 1 when the model has assigned all abnormality attention within the allowed regions for that abnormality, and equals 0 when the model has only assigned attention to forbidden regions. The OrganIoU is calculated using the model's predicted attention $\mathcal{\tilde{A}} \in \mathbb{R}^{M \times H \times D_1 \times D_2}$ and the attention ground truth $\mathcal{G}_{true} \in \{0,1\}^{M \times H \times D_1 \times D_2}$. The predicted attention is binarized with different thresholds and the optimal threshold chosen for each abnormality on the validation set. Define $allowed$ as the sum of all predicted attention values where $\mathcal{G}_{true} = 1$ and $forbidden$ as the sum of all predicted attention values where $\mathcal{G}_{true} = 0$. Then OrganIoU = $\frac{allowed}{allowed+forbidden}$.

\subsection{AxialNet and HiResCAM enable explainable multiple abnormality prediction in CT volumes}

\subsubsection{AxialNet outperforms competing methods}

\begin{table}
\caption{RAD-ChestCT validation set classification performance  and localization performance using the predefined 2,000 train/1,000 val subset \cite{Draelos} for computational feasibility. Classification performance is reported as median AUROC (area under the receiver operating characteristic) while localization performance is reported as mean OrganIoU. The proposed AxialNet architecture outperforms all previously published multilabel CT scan classifiers (CTNet, 3DConv based on 3D convolutions, and BodyConv), as well as a new BodyCAM architecture detailed in the Appendix, and ablated versions of AxialNet. OrganIoU was calculated at the last convolutional layer of all models. No OrganIoU could be calculated for CTNet as the spatial relationship between the output of the last convolutional layer and the input has been disrupted due to convolution over features.}
\label{tab:ALTandABL}
\begin{center}
\footnotesize
\begin{tabular}{llccc}
\multicolumn{5}{c}{Architecture Comparison and Ablation Study} \\
\toprule
                      &                 &                 & \multicolumn{2}{c}{OrganIoU} \\
                                                            \cmidrule(lr){4-5}
    Model             &                     & AUROC          & Grad             & HiRes    \\ \midrule
                & CTNet \cite{Draelos}      & 66.3           & -                & -       \\
Alternative     & 3DConv \cite{Draelos}     & 53.8           & 3.3              & 4.5     \\
Architectures   & BodyConv \cite{Draelos}   & 50.2           & 0.0              & 14.6    \\ 
                & BodyCAM                   & 62.0           & 15.7             & 15.7    \\ \midrule
AxialNet        & MaxPool                   & 65.5           & 9.3              & 10.9   \\
Ablation        & RandInitResNet            & 63.1           & 15.1             &  14.1     \\
Study           & NoCustomConv              & 56.5           & 10.0             & 19.7       \\ \midrule
Proposed        & AxialNet          & \textbf{67.5}  & \textbf{18.7}    & \textbf{21.5}      \\ \bottomrule
\end{tabular}
\end{center}
\end{table}

AxialNet is the first multilabel CT scan abnormality classification model that has built-in explainability. To understand how the built-in explainability affects performance, we compare AxialNet to previously published and alternative architectures, including fully ``black box'' models such as CTNet. We find that AxialNet outperforms these models on both classification and localization (Table \ref{tab:ALTandABL}). For an analysis of abnormality-specific performance, see Appendix Section \ref{sec:abn-specific-perf}.

\subsubsection{AxialNet ablation study}

We additionally performed an ablation study to gain further insight into different components of the AxialNet architecture. The ablation study demonstrates that the following attributes, included in our final proposed AxialNet model, yield improved performance: multiple instance learning based on average pooling (instead of max pooling, denoted as MaxPool in Table \ref{tab:ALTandABL}), pre-training the ResNet on ImageNet (instead of randomly initializing it, denoted as RandInitResNet), and including the custom convolution layers (rather than excluding them, denoted as NoCustomConv).

\subsubsection{AxialNet enables explainable prediction through identification of top slices}

\begin{figure}
    \centering
    \includegraphics[scale=0.6]{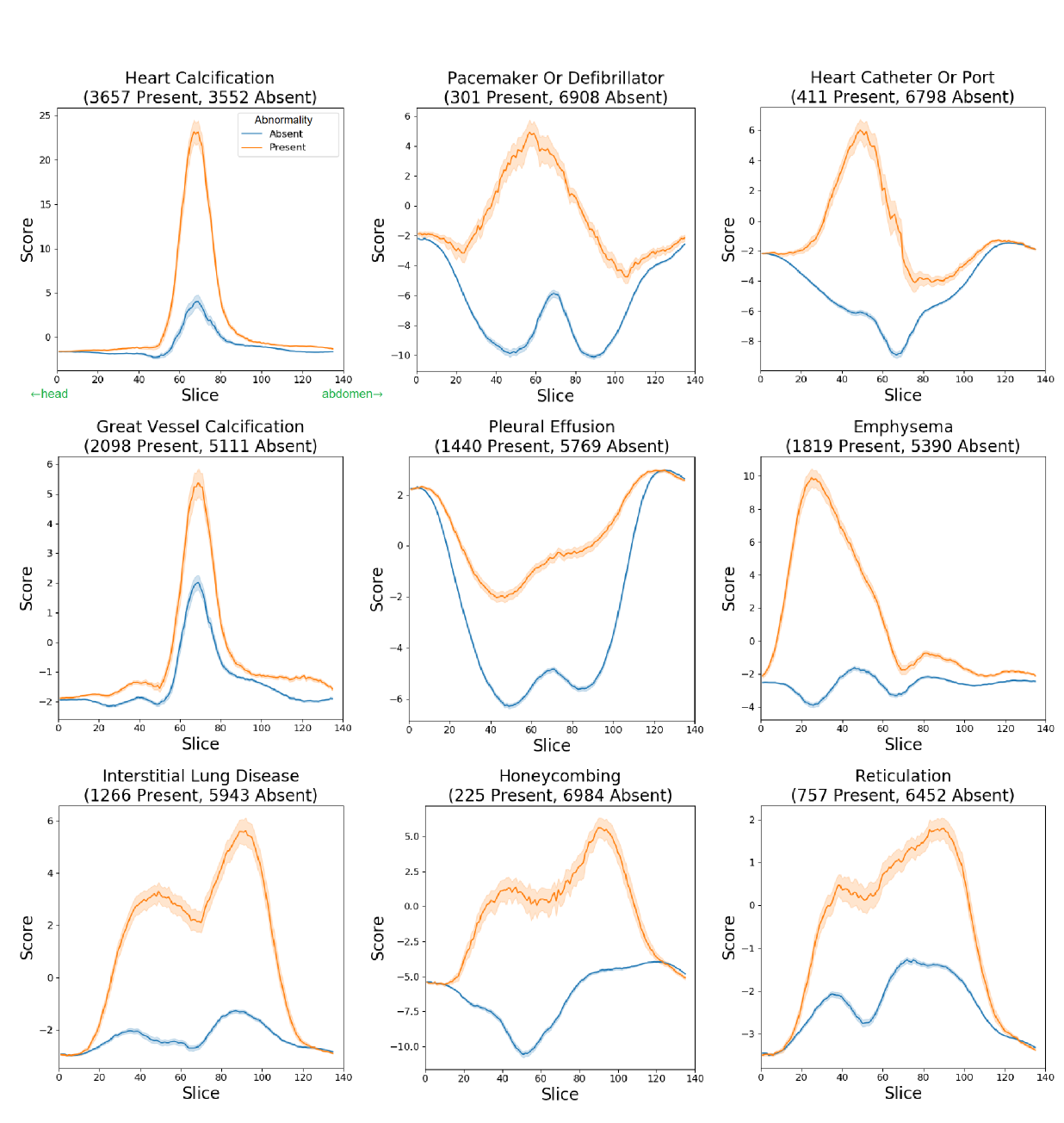}
    \caption{AxialNet provides explainability through per-slice abnormality scores, and the explanations in turn suggest that the model may have learned some medical concepts. This figure provides a summary of the per-slice abnormality scores $\mathbf{C} \in \mathbb{R}^{M \times H}$ across the 7,209 RAD-ChestCT test set CT volumes, for 9 example abnormalities. The matrices $\mathbf{C}$ were calculated for each scan using the final AxialNet $\mathcal{L}_{class}+\lambda \mathcal{L}_{mask}$ whole dataset model. The orange line depicts the mean score and 95\% confidence interval for scans in which the listed abnormality is present, while the blue line depicts the mean score and 95\% confidence interval for the scans in which the abnormality is absent. Axial slice 0 is closer to the head, while slice 135 is closer to the abdomen. The medical concepts demonstrated in this figure are described in detail in the text. Best viewed in color.}
    \label{fig:orangeblueline}
\end{figure}

In addition to outperforming competing methods, the AxialNet model provides explainability via the matrix of all per-slice abnormality scores $\mathbf{C} \in \mathbb{R}^{M \times H}$ (abnormalities $\times$ slices) which is produced for each input CT volume as an intermediate step in the model's computations. The $\mathbf{C}$ matrix quantifies the contribution of each axial slice to each final abnormality score. Figure \ref{fig:orangeblueline} provides a visualization of $\mathbf{C}$, showing per-slice abnormality scores aggregated across all test set volumes, for particular abnormalities of interest. The visualization demonstrates that $\textbf{C}$ displays patterns consistent with medical knowledge. 

First, Figure \ref{fig:orangeblueline} shows a general pattern in which the orange line (for scans containing the abnormality) is higher than the blue line (for scans lacking the abnormality), indicating higher scores when the abnormality is present, as desired. On the top row, example heart abnormalities tend to have peak scores in central slices where the heart is found. Furthermore, the scores for ``heart calcification'' and ``great vessel calcification'' have a similar distribution across slices, which is reasonable since these abnormalities are related - though great vessel calcification scores are comparatively higher in slices 80$+$, which makes sense as the aorta (a great vessel) descends into the abdomen. For ``pleural effusion,'' the model tends to yield high scores towards the lower section of the chest cavity ($\approx$slices 80-120), which is reasonable because pleural effusions frequently collect in the pleural spaces next to the lung bases. The high scores for pleural effusion in the upper lungs may be an artefact of symmetry - the model may have learned to find the lung bases through a relative decrease in the proportion of the slice occupied by lung tissue, and is detecting this relative decrease again at the apices. The ``emphysema'' scores peak towards the upper lobes of the lungs ($\approx$slices 0-50), which is consistent with the fact that the most common form of emphysema (centrilobular emphysema) is typically most visible in the upper lungs \cite{takahashi2008imaging}. Finally, ``interstitial lung disease,'' ``honeycombing,'' and ``reticulation,'' when present, have high predicted scores throughout the entire lung field, which is reasonable as all of these abnormalities tend to be diffuse \cite{sundaram2008accuracy}.

\subsubsection{HiResCAM provides additional explainability}

AxialNet's built in slice-level explanations are useful for gaining insight into the model's behavior. HiResCAM augments these explanations further, by providing insight into sub-slice locations that contribute to a particular abnormality prediction. Figure \ref{fig:heatmapmanyexamples} includes visualizations of HiResCAM explanations for predictions of several different abnormalities. These heat maps narrow down the explanation to a particular part of each slice.

\subsubsection{HiResCAM explanations are faithful to the model, while Grad-CAM produces misleading explanations}

The Grad-CAM explanation method has been used in a variety of medical imaging applications. However, Grad-CAM sometimes highlights locations the model did not use, yielding misleading explanations.

\begin{figure}
    \centering
    \includegraphics[scale=0.1]{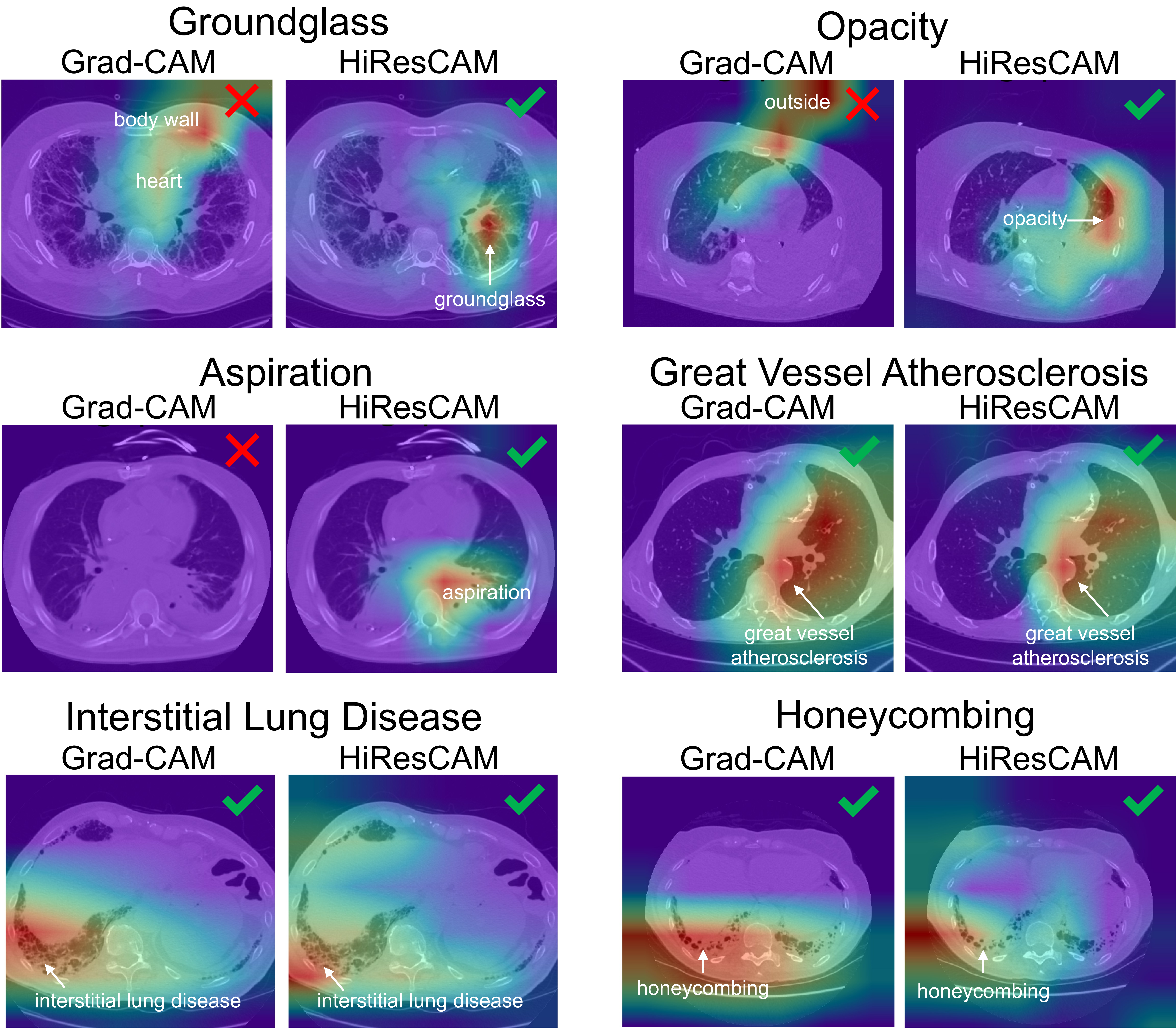}
    \caption{A comparison of Grad-CAM and HiResCAM explanations for true positive predictions of AxialNet $\mathcal{L}_{class}+\lambda \mathcal{L}_{mask}$ trained on the whole RAD-ChestCT data set. For groundglass, opacity, and aspiration, the feature-focused Grad-CAM explanation creates the incorrect impression that the model has made a prediction for a lung abnormality based on an irrelevant anatomical region, while the location-focused HiResCAM explanations highlight the lung locations that the model actually used. For great vessel atherosclerosis, interstitial lung disease, and honeycombing, both Grad-CAM and HiResCAM produced reasonable results. The fact that Grad-CAM can sometimes highlight plausible regions may explain its popularity in medical imaging. Best viewed in color.}
    \label{fig:heatmapmanyexamples}
\end{figure}

We generated HiResCAM and Grad-CAM explanations, holding the model, abnormality, and input CT volume constant, and found that these explanations were often different. Because HiResCAM is mathematically guaranteed to highlight only the locations the model actually used to make a prediction, this means that Grad-CAM is wrong whenever it disagrees with HiResCAM. Figure \ref{fig:heatmapmanyexamples} compares HiResCAM and Grad-CAM explanations across a variety of abnormalities. Sometimes the HiResCAM and Grad-CAM explanations appear similar, \textit{e.g.} for ``great vessel atherosclerosis,'' ``interstitial lung disease,'' and ``honeycombing.'' However, other times Grad-CAM creates the incorrect impression that the model made predictions for lung abnormalities based on the heart or body wall, when in fact the model did rely on the lungs as illustrated in the HiResCAM explanations (``groundglass,'' ``opacity,'' and ``aspiration''). We hypothesize that Grad-CAM focuses on the wrong organ in these examples because these lung abnormalities are ``light grey'' and may activate features that detect this ``light grey'' quality; however, the heart is more ``light grey'' than any adjacent lung tissue, so feature-focused Grad-CAM fixates incorrectly on the heart.

Furthermore, by focusing on locations the model did not actually use, Grad-CAM explanations generally produce worse organ localization of abnormalities, as can be seen by Grad-CAM's lower OrganIoU in Tables \ref{tab:ALTandABL} and \ref{tab:finaltestiou} for the 3DConv, BodyConv, MaxPool, NoCustomConv, and AxialNet models. (Grad-CAM and HiResCAM have identical OrganIoU for BodyCAM because this is a CAM architecture, and Grad-CAM and HiResCAM are alternative generalizations of CAM, meaning they yield identical explanations for CAM architectures only \cite{draelos2020usehirescam}.)

It is interesting to consider HiResCAM and Grad-CAM visualizations for CT scans that contain no abnormalities. This topic is discussed in depth in Appendix Section \ref{sec:normal-CTs-viz}.

\subsection{The mask loss improves model localization of abnormalities} \label{sec:mask-improves-organiou}

Inspection of the explanations for the AxialNet model trained with only $\mathcal{L}_{class}$ revealed that this model sometimes makes predictions using unexpected locations, and therefore may be exploiting some spurious correlations (discussed further in Section \ref{sec:casestudy} and Figure \ref{fig:greatvessel}). The goal of the mask loss is to reduce use of spurious correlations and encourage the model to predict abnormalities from more medically reasonable locations; the latter characteristic can be measured with OrganIoU. OrganIoU is lower when the model exploits spurious correlations in irrelevant locations, and is higher when the model predicts each abnormality based on the organs in which that abnormality appears. This is true only for the OrganIoU calculated using HiResCAM explanations, since HiResCAM explanations are faithful to the model while Grad-CAM explanations are not. We found that the mask loss did improve HiResCAM OrganIoU as desired, by 33\% overall, yielding state-of-the-art abnormality localization performance (Table \ref{tab:finaltestiou}). 

\begin{table}
\caption{Test set results for two models trained on the whole RAD-ChestCT data set of 36,316 CT volumes: AxialNet $\mathcal{L}_{class}$ only, versus AxialNet $\mathcal{L}_{class}+ \lambda \mathcal{L}_{mask}$. The proposed mask loss yields state-of-the-art performance with a 33\% improvement in organ localization of abnormalities (20.7 $\xrightarrow[]{}$ 27.6 OrganIoU). }
\label{tab:finaltestiou}
\begin{center}
\footnotesize
AxialNet Test OrganIoU on Whole RAD-ChestCT Dataset
\begin{tabular}{@{}ccccc@{}}
\toprule
 \multicolumn{2}{c}{AxialNet $\mathcal{L}_{class}$}   & \multicolumn{2}{c}{AxialNet $\mathcal{L}_{class} + \lambda \mathcal{L}_{mask}$ } & \multicolumn{1}{c}{Overall} \\
       \cmidrule(lr){1-2}
        \cmidrule(lr){3-4}
 Grad-CAM   & HiResCAM      & Grad-CAM      & HiResCAM      &   $\uparrow$      \\ \midrule
 20.1       & 20.7           & 21.4          & \textbf{27.6}  & \textbf{+33\%}        \\ \bottomrule
\end{tabular}
\end{center}
\end{table}

\subsection{Great vessel atherosclerosis case study: benefits of HiResCAM and the mask loss} \label{sec:casestudy}

In this section, we provide a case study that illustrates how HiResCAM explanations can reveal when a model is exploiting spurious correlations, and how the mask loss may reduce this undesirable behavior. 

\begin{figure}
    \centering
    \includegraphics[scale=0.1]{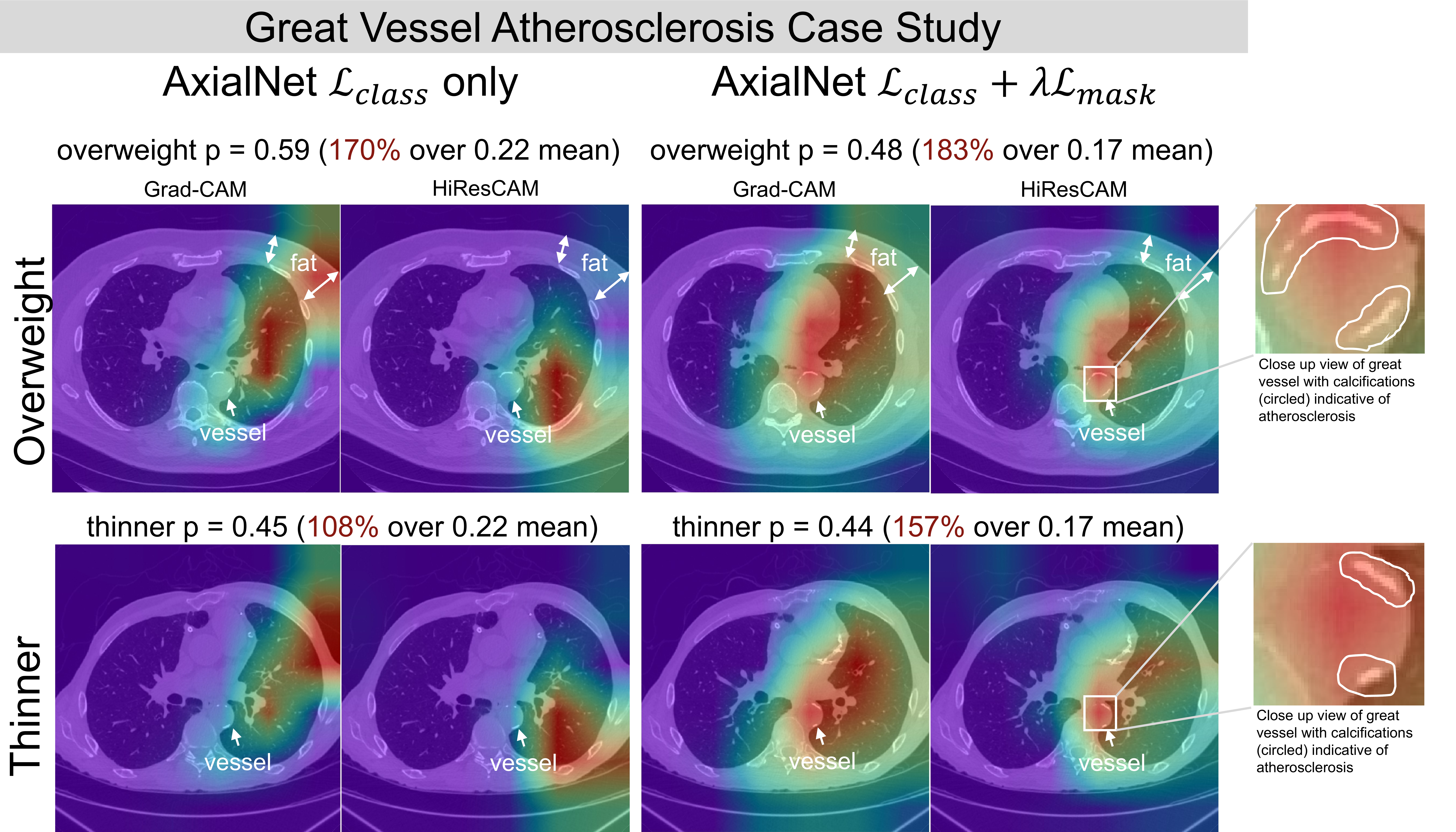}
    \caption{Great vessel atherosclerosis case study in which HiResCAM illustrates that the initial AxialNet $\mathcal{L}_{class}$ model may have exploited the correlation between atherosclerosis and body wall fat, an undesirable behavior that was reduced by using the mask loss. Anatomy annotations added for clarity. Best viewed in color. Further analysis of the figure is provided in the text. Best viewed in color.}
    \label{fig:greatvessel}
\end{figure}

The case study is depicted in Figure \ref{fig:greatvessel}. The left side of Figure \ref{fig:greatvessel} depicts the AxialNet $\mathcal{L}_{class}$ model. While this AxialNet $\mathcal{L}_{class}$ model appropriately predicts high likelihood of atherosclerosis (probability 170\% over the mean) in an overweight patient, the explanation highlights the fat in the body wall and \textit{not} the great vessel, suggesting that the AxialNet $\mathcal{L}_{class}$ model may be inappropriately exploiting the correlation between body fat and atherosclerosis to make its atherosclerosis prediction. High fat in the body wall (\emph{i.e.}, overweight or obesity) is a known risk factor for atherosclerosis \cite{yoo2014adipokines}. However, it is inappropriate to directly use a patient's weight to predict atherosclerosis, because it is possible to be overweight without great vessel atherosclerosis and it is also possible to have great vessel atherosclerosis without being overweight. Great vessel atherosclerosis should be diagnosed by inspecting the great vessel itself for signs of atherosclerosis, which include calcifications like those circled in the inset on the right. Indeed, the AxialNet $\mathcal{L}_{class}$ model's strategy of exploiting the patient's obesity can backfire. The bottom row shows a thinner patient who still has great vessel atherosclerosis in spite of their lower body fat. Here, the AxialNet $\mathcal{L}_{class}$ model again places too much focus on the body wall and produces a score only 108\% above average in spite of the obvious atherosclerosis visible in the vessel.

The right side of Figure \ref{fig:greatvessel} depicts the AxialNet $\mathcal{L}_{class} + \lambda \mathcal{L}_{mask}$ model, which has been trained with the mask loss to encourage predictions from within relevant anatomical regions. This model's explanations for great vessel atherosclerosis appear to focus more on the great vessel itself. The benefit is seen for the thinner patient, where the AxialNet $\mathcal{L}_{class} + \lambda \mathcal{L}_{mask}$ model is able to produce a score 157\% above average in spite of the patient's thinner body habitus. It appears that the mask loss may have discouraged exploiting the body wall for atherosclerosis prediction, and was thus able to yield a model that relied more on actual signs of atherosclerosis.

\section{Discussion}

\subsection{Classification model explanation versus segmentation}

Those familiar with abnormality segmentation models will recognize that pixel-wise abnormality predictions produced by segmentation models are often more precise than the AxialNet explanations produced in our experiments. To put our experimental results in context, in this section we explore the difference between model explanation and segmentation.

First, explanation and segmentation have different goals. A visual explanation should reveal the image regions a classification model used for prediction, even when those regions do not overlap the abnormality of interest - \textit{especially} when those regions do not overlap the abnormality of interest, as that suggests a problem with the model. In contrast, the main goal of segmentation is to identify all pixels that are part of an abnormality as accurately as possible. The main goal of our work is classification model explanation.

Second, explanation and segmentation are based on different information. Explanations are derived from classification models trained only on presence/absence labels - in this case, a label that applies to a whole CT volume. A segmentation model is trained on pixel-level labels; a CT has on the order of $500 \times 500 \times 500 = 1.25 \times 10^8$ pixels. Thus a segmentation model is trained on orders of magnitude more labels than a classification model. The downside to segmentation is the expense of obtaining the pixel-level labels, which limits the number of training examples and the number of abnormalities that can be considered. Still, for the highest quality abnormality localization, a fully supervised segmentation model is the best choice.

Weakly-supervised segmentation has the goal of a segmentation model but uses only the information available to a classification model. For further discussion of the distinction between model explanation and weakly-supervised segmentation, see prior work \cite{draelos2020usehirescam}.

\subsection{Clinical applicability}

The primary clinical relevance of this work is to demonstrate explainability approaches that may increase end-user trust and help prevent deployment of classification models that leverage non-medical confounding factors. The AxialNet $\mathcal{L}_{class} + \lambda \mathcal{L}_{mask}$ model does not presently achieve high enough performance for real-world deployment, but it demonstrates promising properties, including slice-level and sub-slice explanations that appear to incorporate medical information. Prior work has demonstrated substantial performance gains in computer vision tasks through simply increasing model capacity and dataset size \cite{sun2017revisiting}. Our AxialNet implementation was constrained by the GPU memory available in our research machines, and likewise the RAD-ChestCT dataset has a fixed size of 36,316 volumes. It is conceivable that greater model capacity and scaling the training data to millions of scans could yield significant performance gains that might make our proposed approach deployable in the future. Massive increases in dataset size are possible because over 70 million CT scans are obtained each year in the United States alone \cite{bosch2016trends}, and the necessary ground truth labels needed for training can be extracted automatically from associated reports \cite{Draelos,irvin2019chexpert}.

\subsection{Limitations and future directions} 

A limitation of our experiments is that they all relied on the RAD-ChestCT dataset. To the best of our knowledge, there are no other classification datasets available that include tens of thousands of whole CT volumes annotated with over 80 diverse abnormalities. Because the RAD-ChestCT dataset does not include any abnormality segmentation maps, we were also unable to calculate IoU or DICE coefficient, and had to use a less detailed metric, OrganIoU, to estimate organ localization of abnormalities. In the future it would be useful to build on the RAD-ChestCT dataset by manually obtaining ground truth segmentation maps, and thus enable calculation of IoU or DICE coefficient. Before any clinical deployment, it would also be critical to perform external validation and investigate whether models trained on RAD-ChestCT generalize to other datasets that include different protocols, scanners, patient populations, and abnormality distributions.

A limitation of the AxialNet model is its tendency to focus on small discriminative regions, which is a frequent weakness for classification CNNs \cite{lee2019ficklenet, li2018GAIN, chang2020SCCAM, kolesnikov2016SEC}. Consequently, some of AxialNet's explanations highlight only one example of an abnormality, \textit{e.g.} bilateral groundglass highlighted on only one side, or only one nodule highlighted out of many. Exploring methods to encourage AxialNet to leverage all relevant examples without focus spreading to irrelevant organs is thus a promising direction for future work.

A general limitation of visual explanation methods is that these methods show \textit{what} parts of the input are being used but do not provide insight into \textit{how} these parts of the input are used. HiResCAM and other visual explanations suggest a model's behavior by highlighting particular locations, but they do not indicate what functions were computed on these locations. In the worst case scenario, a model could look at the right region for the wrong reasons - for example, detecting pneumonia via post-processing and compression artefacts located within the lung fields - and then the human observer would have no way to identify the model's mistake. Overall, building truly interpretable computer vision models in which both the locations used and the functions computed on them are transparent remains a major unsolved problem.

\section{Conclusions}

In this study, we introduced the new task of explainable multiple abnormality classification in chest CT volumes. We presented a multiple instance learning CNN architecture, AxialNet, that specifies top axial slices contributing most to each abnormality prediction. We proved that the HiResCAM gradient-based explanation method is guaranteed to highlight the regions the AxialNet model used. We then proposed a mask loss that enables AxialNet to achieve better organ localization of abnormalities and thus more medically plausible explanations. Calculation of the mask loss is enabled by PARTITION, the first approach for automatic identification of 3D abnormality-specific allowed regions. Overall, our innovations result in a 33\% improvement in organ localization of abnormalities, and represent the first effort towards explainable multiple abnormality prediction in chest CT volumes.

\section*{Acknowledgements}

The authors would like to thank the Duke Protected Analytics Computing Environment (PACE), particularly Mike Newton and Charley Kneifel, PhD, for providing the computing resources and GPUs needed to complete this work. The authors also thank Paidamoyo Chapfuwa, PhD, for thoughtful comments on a previous version of the manuscript, David Dov, PhD, for discussion of multiple instance learning, and Geoffrey D. Rubin, MD, FACR, for helpful remarks on the explanations.

\section*{Funding Sources}

This work was supported in part by the National Institutes of Health (NIH) Duke Medical Scientist Training Program Training Grant (GM-007171).


\bibliographystyle{unsrt}
\bibliography{Sources}

\section{Appendix}

\subsection{Further motivation for the mask loss} \label{sec:furthermaskloss}

The classification loss $\mathcal{L}_{class}$ determines the extent of attention (\emph{e.g.} a whole lung or only a lobe) in allowed regions, while the mask loss $\mathcal{L}_{mask}$ discourages attention in forbidden regions. 

\textit{Classification loss:} The amount of attention placed within allowed regions, determined by $\mathcal{L}_{class}$, can vary from abnormality to abnormality. The model could attend to the entire allowed region (useful for diffuse abnormalities) or a small sub-part of the allowed region (useful for focal abnormalities). 

\textit{Mask loss}: The mask loss is
\begin{align}
    \mathcal{L}_{mask} =  - \sum_{i: \mathcal{G}_{true} = 0} \log (1-\tilde{a}_i).
\end{align}
where $\tilde{a}_i$ accesses all elements of $\mathcal{\tilde{A}} \in \mathbb{R}^{M \times H \times D_1 \times D_2}$, the model's predicted attention, and $\mathcal{G}_{true} \in \{0,1\}^{M \times H \times D_1 \times D_2}$ defines an allowed region for the attention for each abnormality. 

The mask loss is half of a segmentation cross entropy objective \cite{jadon2020surveyofloss}, specifically the part of the objective where the ground truth is equal to 0. To minimize the mask loss the model must not increase the abnormality score using forbidden regions (where $\mathcal{G}_{true} = 0$), which are regions outside the organ(s) in which that abnormality is found in a particular scan. The portion of the full segmentation objective relating to $\mathcal{G}_{true} = 1$ is excluded because we do not want to force the model to attend to the entire relevant organ. This is because most abnormalities do not occupy the entire organ, especially focal abnormalities such as nodules and masses. 

In practice we find that training with $\mathcal{L}_{mask}$ is stable. We explored an alternative mask loss formulation based on an $L_2$ norm but found that its performance was inferior.

The mask loss is computed in a low-dimensional space, at the level of the last convolutional layer where the HiResCAM explanation is produced. This is orders of magnitude more computationally efficient than computing a mask loss in the input space.

\textit{Loss conceptual example: right lung nodule:} For a nodule in the right lung, the mask loss will discourage the model from predicting this nodule using forbidden regions, \emph{i.e.} anatomy anywhere outside of the right lung. For example the mask loss will discourage the model from exploiting liver nodules to predict the lung nodule (in a patient with metastatic cancer, there may be cancerous nodules and masses in multiple organs, but to predict specifically ``right lung nodule'' only the right lung should be used). The goal of the classification loss is to enable the model to make a correct prediction, so the classification loss should encourage the model to predict the nodule from the small part of the lungs in which it is actually found. Ideally, the final explanation for a right lung nodule should cover the relevant part of the right lung, and nowhere outside of the right lung.

\textit{Features in one organ providing clues for another organ}: It is true that sometimes, features in one organ may provide clues for another organ, such as the case of the metastatic cancer patient described above. We split abnormalities that occur in multiple organs into different labels to encourage the model to learn differences between organs. Developing models to leverage relationships across organs while precisely distinguishing between them is an interesting direction for future work.

\subsection{PARTITION: Creating an attention ground truth} \label{sec:creatingattngrtruthdetails}

The PARTITION approach to create $\mathcal{G}_{true}$, the allowed regions for each abnormality, combines location $\times$ abnormality labels extracted from radiology reports with unsupervised organ segmentation, as shown in Figure 4 of the main paper. This section provides further details on PARTITION's subcomponents. 

\subsubsection{Location and abnormality labels} \label{sec:abn-labels-used}

The first step in creating the attention ground truth $\mathcal{G}_{true}$ is obtaining the location $\times$ abnormality labels. 
The location $\times$ abnormality labels are produced via a simple extension of SARLE, the publicly available rule-based automated label extraction method used to create the RAD-ChestCT abnormality labels \cite{Draelos}. SARLE was introduced and evaluated in prior work \cite{Draelos}. The first step of SARLE is phrase classification, in which each sentence of a free text report is analyzed using a rule-based system to distinguish between ``normal phrases'' (those that describe normality or lack of abnormalities) and ``abnormal phrases'' (those that describe presence of abnormalities). The second step of SARLE is a ``term search'' that searches for abnormality-related vocabulary within the ``abnormal phrases'' to identify which specific abnormalities are present. The handling of negation in the first step of SARLE is highly effective, and when combined with the term search yields an average F-score of 0.976 \cite{Draelos}. SARLE is designed to be easily customizable through the addition of extra vocabulary to the term search. We leverage this customizability by adding location terms to the term search, to identify whether each abnormality is located in the lungs, heart, great vessels, mediastinum, or elsewhere. We also make use of the medical definitions of the abnormalities themselves. Many abnormalities reveal their locations by definition, \textit{e.g.} ``pneumonia'' (lung infection) which can only occur in the lungs or ``cardiomegaly'' (enlarged heart) which can only occur in the heart. Specific examples of some of our extensions to SARLE's term search are shown in Table \ref{tab:sarleExpanded}.

\begin{table*}
\caption{Examples of identifying the location of each abnormality from the radiology reports. In Step 1 (Vanilla SARLE) only abnormal phrases are kept. In Step 2 (Vanilla SARLE) the abnormality is identified with a term search. In Step 3 (our addition), the location is identified with a term search.}
\label{tab:sarleExpanded}
\scriptsize
\begin{tabular}{@{}p{3.0cm}p{2.8cm}p{1.4cm}p{1.4cm}p{3.3cm}@{}}
\toprule
Input Sentence                                                  & Step 1  & Step 2 & Step 3 & Comment                                                                                                                          \\ \midrule
the heart is enlarged without pericardial   effusion            & the heart is enlarged                                         & cardiomegaly                                      & heart                                       & Step 2: ``heart is enlarged'' is one of the synonyms for the abnormality ``cardiomegaly.'' Step 3: the word ``heart'' indicates heart as the location \\ \midrule
there is a nodule in the right upper lobe                       & there is a nodule in the right upper lobe                     & nodule                                            & right lung                                  & Step 3: ``right upper lobe'' indicates ``right lung'' as the location                                                                    \\ \midrule
left pneumonia                                                 & left pneumonia                                               & pneumonia                                         & left lung                                  & Step 3: ``pneumonia'' by definition only affects the lung, so ``left pneumonia''   implies the location ``left lung''                  \\ \midrule
calcifications in the aorta                                     & calcifications in the aorta                                   & calcification                                     & great vessel                                & Step 3: the aorta is a great vessel                                                                                                      \\ \midrule
a calcified granuloma is visible in the   apex of the left lung & a calcified granuloma is visible in the apex of the left lung & calcification, granuloma                          & left lung                                   &                                                                                                                                  \\ \midrule
the catheter tip is visible in the SVC                          & the catheter tip is visible in the SVC                        & catheter or port                                  & great vessel                                & Step 3: the superior vena cava (SVC) is a great vessel                                                                                   \\ \midrule
The lungs are clear    &      &      &     & normal finding; no abnormality labels produced      \\ \midrule
The consolidation has resolved    &     &     &     & normal finding; no abnormality labels produced \\ \bottomrule
\end{tabular}
\end{table*}

Table \ref{tab:conceptgroups} includes a full listing of all the labels used in this study, including 51 right and/or left lung labels and 29 mediastinum labels. Abnormalities were subdivided by location meaning that the right lung and left lung were considered distinctly, thus increasing the total number of labels on which models were trained to 131. Radiologists typically distinguish right from left when describing abnormalities. Furthermore, subdividing abnormalities by location often yields a more medically relevant task. For example, calcification in the lungs is usually caused by calcified nodules or calcified granulomas \cite{khan2010calcifiednodule}, whereas calcification in the aorta is typically due to atherosclerosis \cite{de2014computedaortacalc}. A catheter in the lungs is often a pigtail catheter, \emph{e.g.} to treat a pneumothorax \cite{kuo2013smallpigtail}, while a catheter in the superior vena cava (a great vessel) is a central venous catheter \cite{haygood2011centralvenouscatheter}. 

\begin{table}
    \caption{The 80 abnormality labels used in this study. When subdivided by location, so that the right and left lungs are represented separately, the total number of unique labels increases to 131.}
    \label{tab:conceptgroups}
    \scriptsize
    \centering
    \begin{tabular}{p{0.7in}p{4.0in}} \toprule
        \textbf{Location} & \textbf{Abnormalities} \\ \cmidrule{1-2}
        Right   and/or Left Lung (51) & air   trapping, airspace disease, aspiration, atelectasis, bronchial wall   thickening, bronchiectasis, bronchiolectasis, consolidation, emphysema, fibrosis   (lung), groundglass, interstitial lung disease, lung infection, lung   inflammation, lung scarring, lung scattered nodules or nodes, mucous plugging,   plaque (lung), pleural effusion, pleural thickening, pneumonia, pneumothorax,   pulmonary edema, reticulation, septal thickening, tree in bud, lung resection,   lung transplant, postsurgical (lung), bandlike or linear, honeycombing, lung   calcification, lung cancer, lung cavitation, lung cyst, lung density, lung   granuloma, lung lesion, lung lucency, lung lymphadenopathy, lung mass, lung   nodule, lung nodulegr1cm, lung opacity, lung scattered calcification, lung   soft tissue, chest tube, lung catheter or port, lung clip, lung staple, lung   suture \\
        Mediastinum   (includes Heart and Great Vessels) (29) & cardiomegaly,   heart failure, pericardial effusion, pericardial thickening, CABG, heart   transplant, postsurgical (great vessel), postsurgical (heart), sternotomy, coronary   artery disease, great vessel aneurysm, great vessel atherosclerosis, great vessel calcification, great vessel dilation or ectasia, great vessel scattered calc, heart atherosclerosis, heart calcification, heart scattered calc, mediastinum calcification, mediastinum cancer, mediastinum lymphadenopathy, mediastinum mass, mediastinum nodule, mediastinum opacity, great vessel catheter or port, heart catheter or port, heart stent, heart valve   replacement, pacemaker or defib \\ \bottomrule
    \end{tabular}
\end{table}
\normalsize

\subsubsection{Unsupervised segmentation} \label{sec:unsup-seg}

The second step in creating the attention ground truth $\mathcal{G}_{true}$ is obtaining the segmentation maps for the allowed regions. 

\begin{figure}
    \centering
    \includegraphics[scale=0.1]{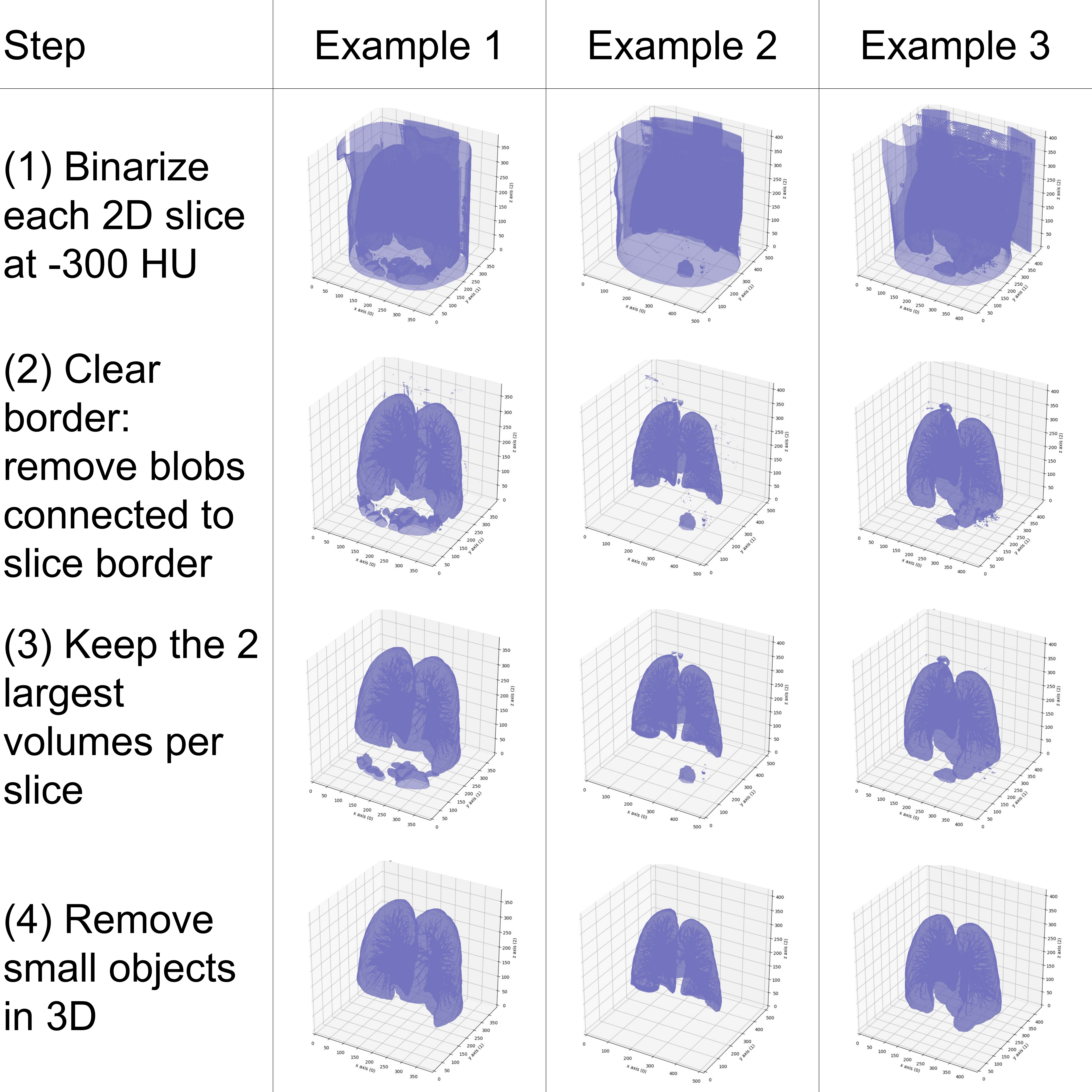}
    \caption{Our unsupervised lung segmentation pipeline using morphological image processing illustrated with 3D renderings at each step, for three different CT volumes. Best viewed in color.}
    \label{fig:segpipeline}
\end{figure}

\textit{Method summary:} Our unsupervised multi-organ segmentation approach includes three stages. In the first stage both lungs are segmented together using morphological image processing, using the four steps shown in Figure \ref{fig:segpipeline}. Next, a bounding box enclosing both lungs is computed and the right lung and left lung are separated via bisection along the midline sagittal plane. Finally, the mediastinum segmentation is defined as the center-left nonlung region inside of the lung bounding box, exploiting the anatomical relationship of the mediastinum and the lungs. The final output is a left lung segmentation, a right lung segmentation, and a mediastinum segmentation that includes the heart and great vessels.

\textit{Evaluation and quality control}. We do not calculate intersection over union (IoU) of our organ segmentations on RAD-ChestCT because there are no ground truth organ segmentations available for RAD-ChestCT. We also do not calculate IoU of our RAD-ChestCT segmentation approach on the CT-ORG or AAPM data sets because of the significant differences in data distribution between RAD-ChestCT and CT-ORG or AAPM discussed in the Related Work section. Like all morphological image processing approaches for lung segmentation, our unsupervised segmentation method is sensitive to threshold values that had to be optimized specifically for RAD-ChestCT. For example the minimum allowed volume size specified in step 4 (removal of small objects) was tuned to 1000000, a value that enables preservation of the right and left lung but elimination of the stomach in RAD-ChestCT. However, this threshold value is inappropriate for any data set that has a different resolution such that a single voxel corresponds to a different physical volume in the real world. Each threshold was chosen in an iterative process that involved manual inspection of hundreds of examples produced with different thresholds. Selecting new customized thresholds for CT-ORG or AAPM would thus not yield an evaluation that is reflective of our algorithm's performance on RAD-ChestCT. A limitation of our work is that we did not manually create segmentation maps to calculate IoU for our unsupervised segmentation pipeline, due to the time consuming nature of creating segmentation ground truth in large 3D volumes. In future work, we would like to manually create this ground truth so that we can quantitatively evaluate our unsupervised segmentation with IoU. For now, we are encouraged by the performance benefit of using the unsupervised segmentation in the mask loss, as well as by the qualitative results for our segmentation approach discussed next.

\begin{figure}
\begin{center}
    \includegraphics[scale=0.18]{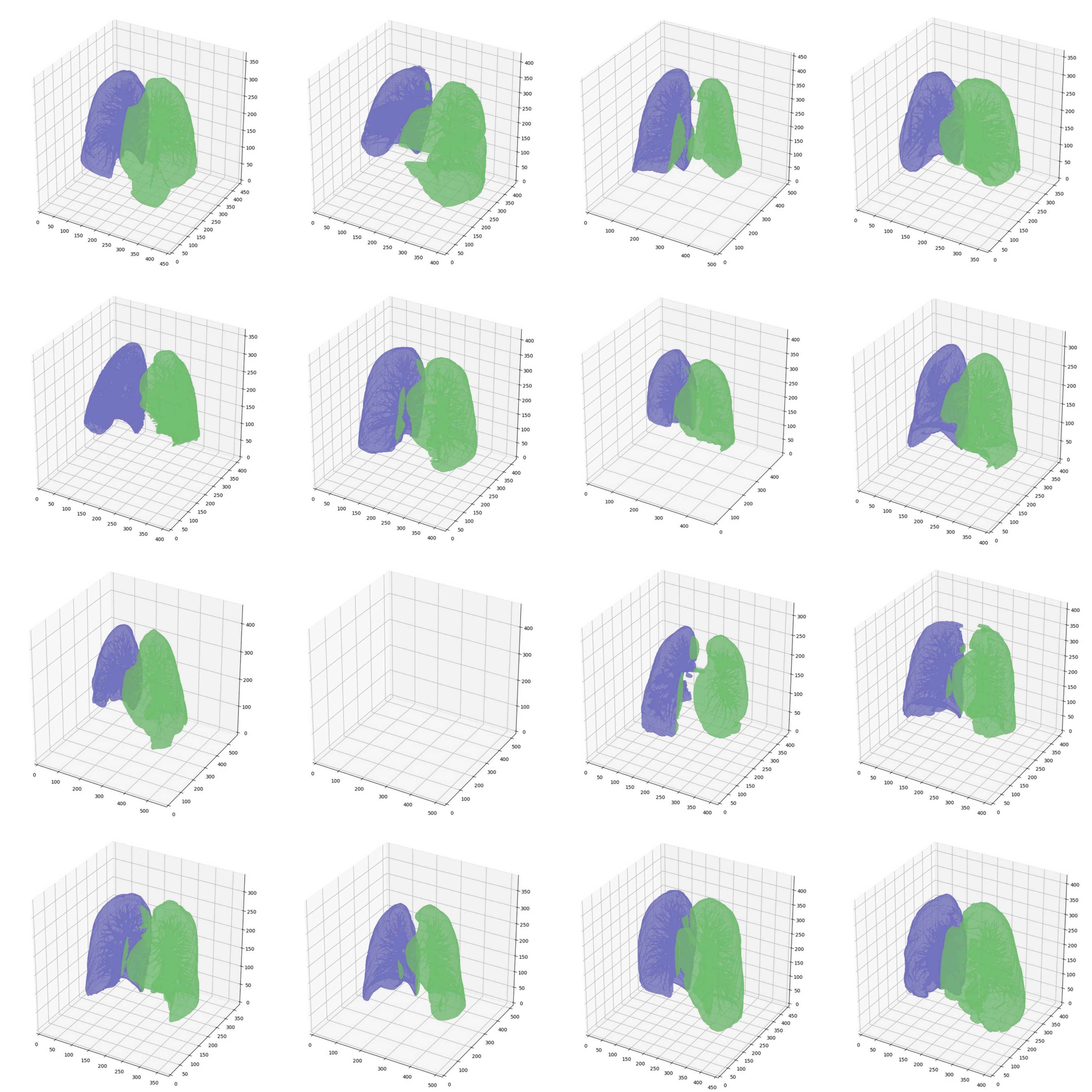}
    \caption{A random selection of full resolution right and left lung segmentations from our unsupervised segmentation method. The scan at row 3, column 2 where both lungs are missing is an example of a scan for which the unsupervised segmentation failed; for this scan a heuristic mask was used in computing the mask loss. Only 4\% of scans require a heuristic mask. Best viewed in color.}
    \label{fig:lungseg}
\end{center}
\end{figure}

\textit{Qualitative results.} We undertook two steps to evaluate our unsupervised segmentation approach. First, we performed manual inspection of numerous randomly-selected segmentations, including 3D renderings and/or 2D projections along the axial, sagittal, and coronal planes. Several randomly-selected right lung and left lung 3D renderings are shown in Figure \ref{fig:lungseg}. Next, we use summary statistics gathered from the training set to assess lung inclusion quality. Analysis of histograms of lung bounding box dimensions paired with visual inspection of outlier scans enabled definition of outlier thresholds that automatically identify when one or both lungs are missing due to severe disease. We quantified the fraction of segmentations in which one or both lungs is missing and found it to be only 4\%, meaning that 96\% of the segmentations pass this basic quality control metric. In creation of the attention ground truth, the 4\% of segmentations that fail quality control are discarded and replaced with a heuristic mask in which the allowed region for the right lung is the right half of the volume, the allowed region for the mediastinum is the center half, and the allowed region for the left lung is the left half. 

\begin{table}
\caption{Validation set median AUROC for AxialNet $\mathcal{L}_{class} + \lambda \mathcal{L}_{mask}$ trained with different organ mask downsampling algorithms and dilation settings. A nearest neighbor downsampling algorithm without dilation achieved the best AUROC and so was used for all other AxialNet $\mathcal{L}_{class} + \lambda \mathcal{L}_{mask}$ experiments including those shown in the main paper. This experiment used the predefined RAD-ChestCT subset of 2,000 training volumes and 1,000 validation volumes \cite{Draelos}.}
\label{tab:downsampling}
\small
\begin{center}
AxialNet Performance with PARTITION Variants
\begin{tabular}{lcc}
\toprule
\textbf{Downsampling Algorithm} & \textbf{Dilation}  & \textbf{AUROC}  \\ \midrule
Area                   & False     & 0.658         \\
                       & True      & 0.662         \\
Trilinear              & False     & 0.661         \\
                       & True      & 0.671         \\
Nearest Neighbor       & False     & \textbf{0.679}\\
                       & True      & 0.654         \\ \bottomrule
\end{tabular}
\normalsize
\end{center}
\end{table}

\subsubsection{PARTITION and mask loss implementation details} \label{sec:downsampling}

In order to calculate the mask loss, the dimensions of the attention ground truth must match the dimensions of the low-dimensional CT representation. Thus, the raw segmentation masks were downsampled to $\{0,1\}^{H \times D_1 \times D_2}$, which has the additional side benefit of ``smoothing out'' any small errors in the segmentation masks. We compared nearest neighbors, trilinear, and area downsampling algorithms, with and without morphological dilation of the downsampled mask. Morphological dilation produces more expansive allowed regions and is thus a more permissive approach. The nearest neighbors algorithm with no dilation yielded the best performance as shown in Table \ref{tab:downsampling}, so this setting was selected to create the attention ground truth used in the final mask loss implementation. 

For all models, the OrganIoU was calculated directly between the predicted attention $\mathcal{\tilde{A}}$ and the attention ground truth $\mathcal{G}_{true}$. We explored calculation of OrganIoU in the input space using upsampling of the predicted attention but found this to be prohibitively computationally expensive (estimated $>$3 weeks of runtime per model).

To improve model training time, the attention ground truth was computed in default axial orientation during the first epoch and then loaded from disk for all subsequent epochs. The data augmentation transformations randomly applied to each CT scan in each epoch were applied dynamically to each sub-volume of that scan's attention ground truth before the mask loss calculation.

\subsection{The BodyCAM architecture and an advantage of AxialNet} \label{sec:bodycam}

Table \ref{tab:ALTandABL} includes performance of a BodyCAM model. We developed the BodyCAM architecture to begin exploring how much value HiResCAM adds over CAM by loosening CAM's architecture requirements. CAM requires a CNN to end in global average pooling followed by one fully connected layer. BodyCAM is essentially a variant of AxialNet that has been modified to meet CAM's architecture requirements. 

The first part of the BodyCAM model is the same as AxialNet, consisting of a 2D ResNet and custom convolutional layers applied to each slice. However, after the custom convolutional layers, BodyCAM uses global average pooling over the $[H,D_1,D_2]$ spatial dimensions, and then a fully connected layer produces the final predictions, whereas AxialNet follows a multiple instance learning setup with a per-slice fully connected layer that produces per-slice predictions which are then averaged to yield the whole-volume predictions.

We hypothesized that AxialNet would outperform BodyCAM, because AxialNet does not include global average pooling and thus can preserve the information about spatial locations of certain features. Our experimental results support this hypothesis, with AxialNet achieving better AUROC than BodyCAM as shown in Table 1 of the main paper. Because BodyCAM follows the CAM architecture, Grad-CAM and HiResCAM produce identical explanations on BodyCAM models, and thus yield the same OrganIoU in Table 1. 

We are encouraged by the result that AxialNet outperforms BodyCAM. While we recognize that the global average pooling step provides the convenient property of enabling input images of variable spatial dimensions, we also believe that if eliminating global average pooling can provide better performance in some tasks then it is a direction worth exploring further. In future work, it would be interesting to systematically investigate the effect of removing the global average pooling step across a variety of architectures and imaging applications.

\subsection{Literature review: how HiResCAM could benefit medical imaging explainability} \label{sec:hirescambenefitlitreview}


This section includes examples of various studies that apply deep learning to medical imaging tasks to illustrate how HiResCAM could positive impact explainability. Table \ref{tab:medicalimaginglitreview} includes examples of previously published studies that apply machine learning to a medical imaging task. The table considers what architecture was used as well as any visual explanation methods that were applied, and analyzes whether there was a risk of a visual explanation method providing the false impression that the model had highlighted the wrong location. The top half of the table includes studies that already incorporated visual explanation methods. The bottom half of the table includes studies that did not report results of a visual explanation method, but which use CNN models for a medical imaging classification task and thus could have had a visual explanation method applied.

It appears that presently, there is risk in medical imaging of relying on faulty visual explanations that highlight locations the model did not use, due to the popularity of Grad-CAM as a visual explanation method and the popularity of custom CNN architectures that end in more than one fully connected layer. In these studies, there are no experiments justifying the need for more than one fully connected layer. Most likely, the decision to include more than one fully connected layer was due to two factors: (a) the authors are likely taking inspiration from architectures like VGG or AlexNet which include multiple fully connected layers; and (b) to the best of our knowledge, no work until recently \cite{draelos2020usehirescam} conveyed the potential explainability benefits of using only one fully connected layer at the end of a CNN, so there may not have been any motivation to limit the number of fully connected layers to one.

In all studies with high risk of wrong explanations, we recommend reducing the number of fully connected layers to one and using HiResCAM as the explanation approach to ensure that the visual explanations reflect the regions the model is using to make predictions.\footnote{If keeping the number of layers the same is a priority, then a reduction in the number of fully connected layers could be accompanied by the addition of an equivalent number of convolutional layers before the final fully connected layer. If it is important not to further shrink the spatial dimensions of the representation, $1 \times 1$ convolutions could be used in these layers.}

\pagebreak
\tiny
\begin{longtable}{p{1.0in}p{0.8in}p{0.8in}p{1.0in}p{0.8in}}
    \caption{Example studies in medical imaging, particularly CT scan analysis, and assessment of whether there was a risk of a visual explanation method highlighting locations the model did not actually use for prediction. When there is a risk (final column orange), that risk could be removed by using HiResCAM for explanations and modifying the architecture to end in only one fully connected layer (``FC layer'').} \\
    \label{tab:medicalimaginglitreview} \\
\toprule
\textbf{Paper} & \textbf{Architecture} & \textbf{Is the architecture a “CAM   architecture”?} & \textbf{Were results of a visual explanation method reported in the paper? }& \textbf{If a visual explanation method were applied, could it highlight incorrect locations?} \\ \cmidrule{1-5}
An explainable deep-learning algorithm for the detection of acute intracranial haemorrhage   from small datasets. \cite{lee2019explainable} & VGG16, ResNet-50,   Inception-v3 and Inception-ResNet-v2 & No for VGG-16. Yes for   ResNet-50, Inception-v3, and Inception-ResNet-v2. & Grad-CAM (paper described   it verbally as “CAM” but the equation shown and references indicate it was Grad-CAM) & \textcolor{orange}{Yes, at a minimum for the   VGG-16 model which ends in $>$1 FC layer.} \\ \cmidrule{1-5}
Comparison of deep   learning approaches for multi-label chest x-ray classification. \cite{baltruschat2019comparison} & ResNet-50 & Yes & Grad-CAM was applied at the   final convolutional layer of an architecture that follows the “CAM   architecture”– thus, this attention approach is mathematically equivalent to CAM & \textcolor{blue}{No} \\ \cmidrule{1-5}
Efficient deep network architectures for fast chest x-ray tuberculosis screening and visualization.  \cite{pasa2019efficient} & Custom CNN ending in global average pooling then one FC layer & Yes & Saliency maps and   Grad-CAM were applied at different layers. From the figure appearance, it appears that Guided Grad-CAM was used rather than vanilla Grad-CAM. & \textcolor{orange}{Yes, because Grad-CAM was  applied at different layers, and also because Guided Grad-CAM appears to have  been used.}  \\ \cmidrule{1-5}
Deep learning to assess  long-term mortality from chest radiographs. \cite{lu2019deep} & Modified Inception-v4 architecture ending in $>1$ FC layer & No & Grad-CAM & \textcolor{orange}{Yes, because Grad-CAM was applied to a CNN ending in $>$1 FC layer.} \\ \cmidrule{1-5}
Classification of Interstitial Lung Abnormality Patterns with an Ensemble of Deep Convolutional   Neural Networks \cite{bermejo2020classification} & Ensemble of 7 CNNs. Architectures include BCNN, MSTAGE-CNN, and MCONTEXT, which all end in 2 or   more FC layers. & No & No & \textcolor{orange}{Yes: architectures ending   in $>$1 FC layers were used.} \\ \cmidrule{1-5}
Weakly-supervised deep   learning of interstitial lung disease types on CT images \cite{wang2019weakly} & Custom CNN ending in   $>$1 FC layer. & No & No & \textcolor{orange}{Yes: an architecture   ending in $>$1 FC layer was used.} \\ \cmidrule{1-5}
Holistic classification   of CT attenuation patterns for interstitial lung diseases via deep   convolutional neural networks \cite{gao2018holistic} & Custom CNN ending in 3 FC   layers. & No & No & \textcolor{orange}{Yes: an architecture ending in $>$1 FC layer was used.} \\ \cmidrule{1-5}
Deep learning for   classifying fibrotic lung disease on high-resolution computed tomography: a   case-cohort study \cite{walsh2018deep} & Inception-ResNet-v2 & Yes & No & \textcolor{blue}{No, if CAM or Grad-CAM or HiResCAM was applied at the last convolutional layer.} \\ \cmidrule{1-5}
Multisource Transfer   Learning With Convolutional Neural Networks for Lung Pattern Analysis   \cite{christodoulidis2016multisource} & Custom CNN ending in 3 FC   layers. & No & No & \textcolor{orange}{Yes: an architecture   ending in $>$1 FC layer was used} . \\ \cmidrule{1-5}
Lung Pattern   Classification for Interstitial Lung Diseases Using a Deep Convolutional Neural Network \cite{anthimopoulos2016lung} & Custom CNN ending in   $>$1 FC layer & No & No & \textcolor{orange}{Yes: an architecture ending   in $>$1 FC layer was used.} \\ \cmidrule{1-5}
Multi-label Deep   Regression and Unordered Pooling for Holistic Interstitial Lung Disease Pattern   Detection \cite{gao2016multi} & AlexNet variant ending   in 2 FC layers. & No & No & \textcolor{orange}{Yes: an architecture ending in $>$1 FC layer was used.} \\ \bottomrule
\end{longtable}
\normalsize



\subsection{Custom convolutional layers}

Code to replicate this paper's experiments will be made publicly available at \href{https://github.com/rachellea}{https://github.com/rachellea}. In the meantime, to clarify the setup of the custom convolutional layers used in AxialNet, we include the PyTorch implementation of these layers here:

\scriptsize
\begin{verbatim}
import torch.nn as nn

def custom_conv_layers():
    return nn.Sequential(
        nn.Conv2d(512, 64, kernel_size = (3,3), stride=(1,1), padding=0),
        nn.ReLU(inplace=True),
          
        nn.Conv2d(64, 32, kernel_size = (3,3), stride=(1,1), padding=0),
        nn.ReLU(inplace=True),
            
        nn.Conv2d(32, 16, kernel_size = (3,3), stride=(1,1), padding=0),
        nn.ReLU(inplace=True),
            
        nn.Conv2d(16, 16, kernel_size = (3,3), stride=(1,1), padding=0),
        nn.ReLU(inplace=True))
\end{verbatim}
\normalsize


\subsection{Abnormality-specific performance} \label{sec:abn-specific-perf}

\begin{table}[]
\caption{Abnormality-specific performance: the RAD-ChestCT test set AUROC with 95\% confidence intervals for the AxialNet model trained on the whole RAD-ChestCT dataset with and without the mask loss, for nine abnormalities. $p$-values comparing the AUROCs were calculated using the DeLong test \cite{delong1988comparing}, with false discovery rate correction for multiple testing \cite{benjamini1995controlling}. The Pos \% column indicates the percent of positive examples in the test set.}
\label{tab:abn-perf}
\centering
\scriptsize
\begin{tabular}{@{}lcccccc@{}}
\toprule
 & & \multicolumn{2}{c}{AxialNet  $\mathcal{L}_{class}$} & \multicolumn{2}{c}{AxialNet $\mathcal{L}_{class} + \lambda \mathcal{L}_{mask}$} & \\ \cmidrule{3-6}
Abnormality & Pos \% & AUROC & 95\% CI & AUROC & 95\% CI & $p$-value\\ \midrule
Lung   nodule & 70.9 & 68.77 & 67.38   - 70.16 & 67.66 & 66.25   - 69.07 &  1.65 $\times 10^{-4}$ \\
Lung   opacity & 47.7 & 62.17 & 60.88   - 63.46 & 58.16 & 56.83   - 59.48 & 2.12 $\times 10^{-12}$ \\
Atelectasis & 28.9 & 70.81 & 69.48   - 72.14 & 71.09 & 69.76   - 72.42 & 0.62 \\
Pleural   effusion & 20.0 & 91.29 & 90.47   - 92.11 & 90.11 & 89.14   - 91.09 & 4.82 $\times 10^{-4}$ \\
Lung consolidation & 14.4 & 73.76 & 72.12   - 75.40 & 67.76 & 65.89   - 69.62 & 1.98 $\times 10^{-15}$ \\
Lung   mass & 7.9 & 69.86 & 67.44   - 72.28 & 65.36 & 62.98   - 67.73 & 3.42 $\times 10^{-6}$ \\
Pericardial   effusion & 15.7 & 67.38 & 65.58   - 69.19 & 67.81 & 66.05   - 69.58 & 0.62 \\
Cardiomegaly & 9.6 & 81.41 & 79.75   - 83.08 & 81.36 & 79.69   - 83.02 & 0.87 \\
Pneumothorax & 2.9 & 81.62 & 78.75   - 84.49 & 81.34 & 78.28   - 84.39 & 0.80 \\ \bottomrule
\end{tabular}
\end{table}
\normalsize


Table \ref{tab:abn-perf} shows the performance of the AxialNet model for nine commonly reported abnormalities. These results demonstrate a trend towards the AxialNet $\mathcal{L}_{class}$ model achieving higher classification performance than the AxialNet $\mathcal{L}_{class} + \lambda \mathcal{L}_{mask}$ model, although the difference is not always statistically significant, as indicated by the DeLong test $p$-values. Overall, across all 80 abnormalities, the AUROC (median $\pm$ interquartile range) was 70.9 $\pm$ 21.0 for AxialNet $\mathcal{L}_{class}$ and 68.2 $\pm$ 23.0 for AxialNet $\mathcal{L}_{class} + \lambda \mathcal{L}_{mask}$. We theorize that there are two potential causes for the mask loss degrading classification performance: implementation-specific and inherent.

\textit{Implementation-specific}: the potential implementation-specific reason centers on the organ segmentation step of PARTITION. The organ segmentations are used to generate abnormality-specific allowed regions upon which the mask loss depends. Table \ref{tab:abn-perf} demonstrates that the abnormalities with the most significantly worse mask loss performance (lung opacity, pleural effusion, lung consolidation, and lung mass) are also abnormalities for which the lung segmentation performance is most likely to deteriorate. This is because the lung segmentation step of PARTITION depends on morphological image processing which assumes lung pixels are black, but in these lung abnormalities, some lung pixels become white. If these white pixels are located at the periphery of the lungs it may result in exclusion of these pixels from the final lung segmentation mask. We consider three scenarios:
\begin{itemize}
    \item Errors of only a few pixels will have minimal effect, since the mask loss is calculated in a low-dimensional space using downsampled segmentation masks (Section \ref{sec:downsampling}).
    \item Errors of an extremely large number of pixels will also have minimal effect, since these errors will be detected by our quality control and a heuristic mask used instead (Section \ref{sec:unsup-seg}).
    \item Errors of a moderate number of pixels have the potential to be most damaging, as these could affect the downsampled segmentation mask without triggering use of a heuristic mask. These errors may be causing the mask loss model to have slightly lower classification performance.
\end{itemize}
This potential limitation can be addressed in future work by improving the lung segmentation step. Specifically, deep-learning-based segmentation models trained specifically on manually annotated diseased lungs may result in improved performance of mask loss models.

\textit{Inherent}: We also hypothesize that the mask loss may inherently decrease classification performance. Section \ref{sec:mask-improves-organiou} shows that the mask loss leads to notably better organ localization of abnormalities, as desired, which may make classification more difficult by limiting reliance on correlated abnormalities (\textit{e.g.}, masses in other organs) or scanner artefacts (\textit{e.g.} scanner compression or post-processing features that correlate with abnormalities due to different clinical services using different scanners). Roughly speaking, it is conceivable that the AxialNet $\mathcal{L}_{class}$ model may be achieving higher classification performance than the $\mathcal{L}_{class} + \lambda \mathcal{L}_{mask}$ model by ``cheating.'' Regardless of the true underlying reason, a limitation of the mask loss is a modest decrease in classification performance for some abnormalities. Addressing this limitation is a direction of future research.

\subsection{HiResCAM and Grad-CAM visualizations of CT scans containing no abnormalities} \label{sec:normal-CTs-viz}

We define a ``normal'' CT scan as one lacking all annotated abnormalities, \textit{i.e.} a CT scan for which the ground truth vector consists entirely of zeros. For a normal CT scan processed by a CNN ending in one fully-connected layer (Section \ref{sec:hires-prove}), a HiResCAM visualization is still interpreted as highlighting the regions of the scan that most increase the abnormality score for a selected abnormality. However, it is important not to assume that red color indicates that the model predicts the abnormality is present, due to the ReLU and normalization step.

HiResCAM, like Grad-CAM, includes a final ReLU and normalization step that converts negative values to zero before linearly scaling into the range [0,1]. In this way, the resulting visualization conveys regions that increase the abnormality score while leveraging the full color spectrum \cite{draelos2020usehirescam}. If the model correctly predicted the absence of an abnormality (true negative), then the \textit{un-processed} HiResCAM heatmap will contain negative-trending values that do not yield a sufficiently high abnormality score to predict presence. If the model incorrectly predicted that the abnormality was present (false positive), then the \textit{un-processed} heatmap will contain positive-trending values that do yield a high enough abnormality score to predict presence. In either case, applying the ReLU and normalization steps then shift all values in a heatmap into the [0,1] range so that whatever values are \textit{relatively} more positive are shown in red, and whatever values are \textit{relatively} more negative are shown in purple, even if those values are not literally positive or negative respectively. Here is a toy example with a four-pixel heatmap:
\begin{itemize}
    \item True negative: $normalize(ReLU([[-3,1],[-4,1]])) = [[0,1],[0,1]]$ 
    \item False positive: $normalize(ReLU([[2,-1],[4,2]])) = [[0.5,0],[1,0.5]]$
\end{itemize}

\begin{figure}
    \centering
    \includegraphics[scale=0.4]{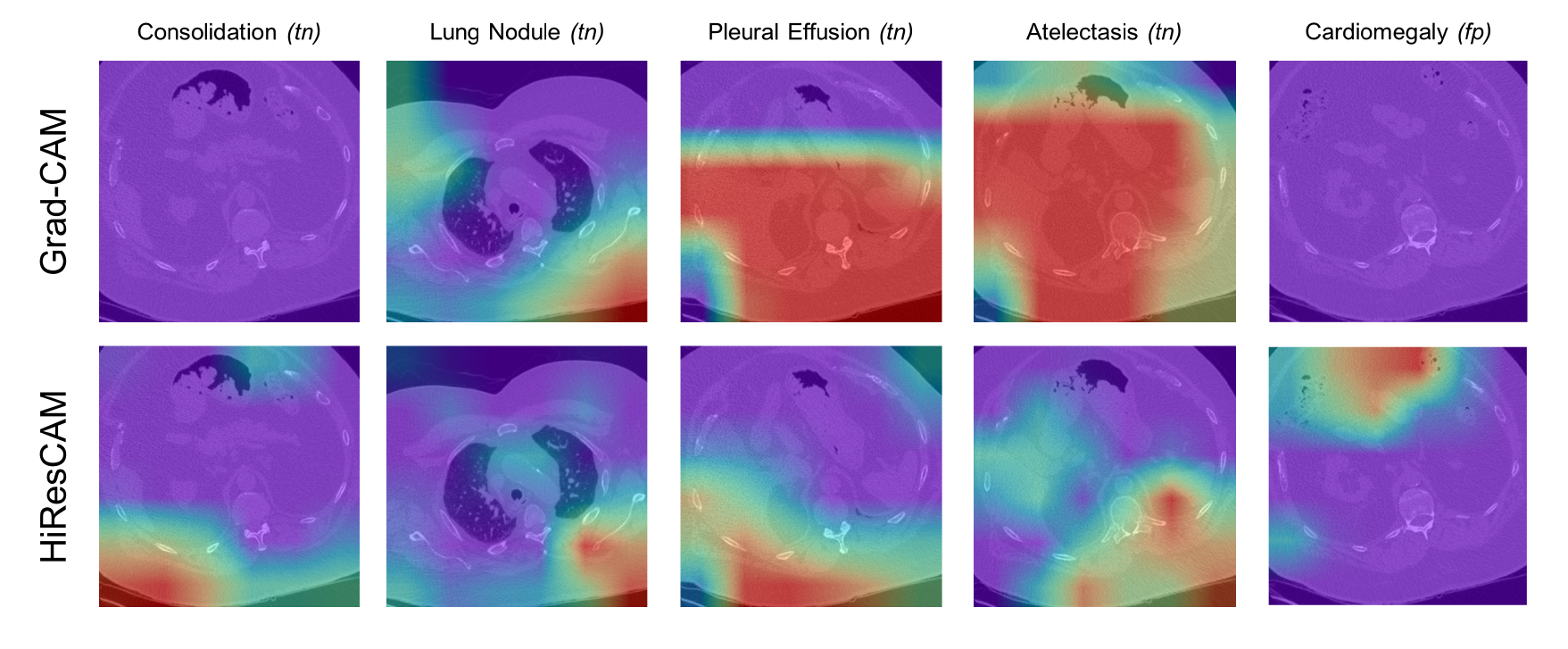}
    \caption{Examples of Grad-CAM and HiResCAM explanations for the AxialNet $\mathcal{L}_{class}+\lambda \mathcal{L}_{mask}$ model, for a CT volume with all abnormalities absent. ``\textit{(tn)}'' indicates the model's prediction was a true negative, while ``\textit{(fp)}'' indicates a false positive.}
    \label{fig:normal-maps}
\end{figure}

Empirically, there does not appear to be much pattern in the visualizations for a normal CT (Figure \ref{fig:normal-maps}), although in theory it is possible that for some normal CTs, HiResCAM could identify areas that are not abnormalities but are ``abnormality-like.'' If the definition of HiResCAM were altered to remove the ReLU and normalization step and apply a universal color scheme across all CT scans, then a true negative prediction would yield a mostly purple HiResCAM visualization, and a false positive prediction would yield a mostly red HiResCAM visualization.

It is not possible to explain what a Grad-CAM visualization means in the case of a normal CT scan, because Grad-CAM explanations do not have guaranteed properties and the faithfulness guarantee does not hold \cite{draelos2020usehirescam}. Figure \ref{fig:normal-maps} does include Grad-CAM visualizations for the same model, input scan, and abnormality, which once again illustrate how Grad-CAM visualizations can be misleading and highlight regions that do not increase the abnormality score.

\end{document}